\documentclass[traditabstract,final]{aa}
\usepackage{graphicx}
\usepackage[small]{caption}
\usepackage{a4wide}
\usepackage{epsfig}
\usepackage{natbib}
\usepackage{fancyhdr}
\usepackage{amsmath}
\usepackage{amssymb}
\usepackage{makeidx}
\usepackage{titletoc}
\usepackage[]{subfigure}
\usepackage{lscape}
\usepackage{hyperref}
\usepackage{wasysym}
\usepackage{wrapfig}
\usepackage{longtable}
\usepackage{aas_macros}
\usepackage{txfonts}
\bibpunct{(}{)}{;}{a}{}{,} 

\begin{document}

\title{The Star Formation \& Chemical Evolution History of the Sculptor Dwarf Spheroidal Galaxy}
\titlerunning{Star Formation History of the Sculptor dSph}

   \author{T.J.L. de Boer\inst{1} \fnmsep \thanks{Visiting astronomer, Cerro Tololo Inter-American Observatory, National Optical Astronomy Observatory, which are operated by the Association of Universities for Research in Astronomy, under contract with the National Science Foundation.}
          \and
          E. Tolstoy\inst{1} \fnmsep \footnotemark[1]
          \and
          V. Hill\inst{2}
          \and
          A. Saha\inst{3} \fnmsep \footnotemark[1]
          \and
          K. Olsen\inst{3} \fnmsep \footnotemark[1]
          \and
          E. Starkenburg\inst{1}
          \and
          B. Lemasle\inst{1}
          \and
          M.J. Irwin\inst{4} 
          \and
          G. Battaglia\inst{5} 
  }
   \offprints{T.J.L. de Boer}

   \institute{Kapteyn Astronomical Institute, University of Groningen,
              Postbus 800, 9700 AV Groningen, The Netherlands\\
              \email{deboer@astro.rug.nl}
             \and
              Universit\'{e} de Nice Sophia-Antipolis, CNRS, Observatoire de la C\^{o}te d$'$Azur, Laboratoire Cassiop\'{e}e, 06304 Nice Cedex 4, France 
             \and
              National Optical Astronomy Observatory\thanks{The National Optical Astronomy Observatory is operated by AURA, Inc., under cooperative agreement with the National Science Foundation.},
              P.O. box 26732, Tucson, AZ 85726, USA
             \and
              Institute of Astronomy, University of Cambridge, Madingley Road, Cambridge, CB3 0HA, UK
             \and
             INAF~$-$~Osservatorio Astronomico di Bologna Via Ranzani 1, I$-$40127, Bologna, Italy
             }

   \date{Received ...; accepted ...}

\abstract{We have combined deep photometry in the B,V and I bands from CTIO/MOSAIC of the Sculptor dwarf spheroidal galaxy, going down to the oldest Main Sequence Turn-Offs, with spectroscopic metallicity distributions of Red Giant Branch stars. This allows us to obtain the most detailed and complete Star Formation History to date, as well as an accurate timescale for chemical enrichment. \\
The Star Formation History shows that Sculptor is dominated by old~($>$10 Gyr), metal-poor stars, but that younger, more metal-rich populations are also present. Using Star Formation Histories determined at different radii from the centre we show that Sculptor formed stars with an increasing central concentration with time. The old, metal-poor populations are present at all radii, while more metal-rich, younger stars are more centrally concentrated. We find that within an elliptical radius of 1 degree, or 1.5 kpc from the centre, a total mass in stars of 7.8$\times$10$^{6}$ M$_{\odot}$ was formed, between 14 and 7 Gyr ago, with a peak at 13$-$14 Gyr ago. \\
We use the detailed Star Formation History to determine age estimates for individual Red Giant Branch stars with high resolution spectroscopic abundances. Thus, for the first time, we can directly determine detailed timescales for the evolution of individual chemical elements. We find that the trends in alpha-elements match what is expected from an extended, relatively uninterrupted period of star formation continuing for 6$-$7 Gyr. The knee in the alpha-element distribution occurs at an age of 10.9$\pm$1Gyr, suggesting that SNe Ia enrichment began $\approx$2$\pm$1Gyr after the start of star formation in Sculptor.}

 \keywords{Galaxies: dwarf -- Galaxies: evolution -- Galaxies: stellar content -- Galaxies: Local Group -- Stars: C-M diagrams}

\maketitle

\section{Introduction}
\label{introduction}
The Sculptor dwarf spheroidal galaxy~(dSph) is a relatively faint~(M$_{V}$$\approx$$-$11.2), well studied system in the Local Group. A distance of~86$\pm$5 kpc has been determined from RR Lyrae star measurements, in good agreement with other distance determinations, such as the tip of the RGB and horizontal branch level~\citep[][and references therein]{Pietrzynski08}. Sculptor is located at high galactic latitude~(b=$-$83$^{\circ}$) with relatively low reddening, E(B$-$V)=0.018~\citep{Schlegel98}. \\
Past studies of the Star Formation History~(SFH) of Sculptor have been made using colour-magnitude diagrams~(CMDs) of varying depth, and spatial coverage. Using CMDs down to the Main Sequence Turn-Off~(MSTO) in a field just outside the core radius~\citet{DaCosta84} determined an age range for Sculptor of 13$\pm$2 Gyr. A small field of view~($\sim$2$^{\prime}$), deep HST image of a field well outside the centre of Sculptor confirmed the ancient age~(15$\pm$2 Gyr) of the bulk of the stars in Sculptor using synthesis CMD analysis, but also showed a small tail of star formation reaching down to more recent times~\citep{Monkiewicz99,Dolphin02}. An effort was made to combine detailed spectroscopic abundances of 5 stars with photometric ages, which also suggested that Sculptor is predominantly old, with a small tail of intermediate age stars~\citep{Tolstoy03}. \\
In a previous qualitative study of wide-field CMDs, the Horizontal Branch~(HB) morphology was found to change significantly with radius~\citep{Majewski99, HurleyKeller99}, which was later linked to a metallicity gradient~\citep{Tolstoy04}. From deep wide-field CMDs reaching the oldest MSTOs, covering a large fraction of Sculptor, this metallicity gradient was also found to be linked to an age gradient~\citep{deBoer2011A}. \\
In Sculptor, wide-field medium resolution \ion{Ca}{ii} triplet spectroscopy of a large number of RGB stars is available~\citep{Battaglia07,Battaglia082,Starkenburg10}, giving a well defined spectroscopic Metallicity Distribution Function~(MDF) for stars with ages $\ge$1.5 Gyr old. In the central 25$^{\prime}$ diameter region of Sculptor high resolution~(HR) spectroscopy~\citep[Hill et al., in prep, see][]{Tolstoy09} of stars on the upper RGB provides detailed abundances of $\alpha$-elements~(O, Mg, Ca, Si, Ti) as well as r- and s-process elements~(Y, La, Ba, Eu, Nd). Furthermore, within the central 15$^{\prime}$ diameter region stars have been observed using medium resolution spectroscopy going down to fainter magnitudes~\citep{Kirby09,Kirby10}, giving [Fe/H] as well as $\alpha$-element abundances. 
\begin{figure}[!ht]
\centering
\includegraphics[angle=0, width=0.5\textwidth]{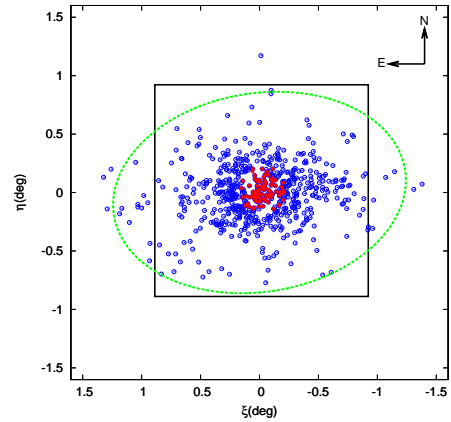}
\caption{Coverage of the photometric and spectroscopic observations across the Sculptor dwarf spheroidal galaxy. The solid~(black) square denotes the full coverage of the CTIO 4m/MOSAIC fields. The~(blue) open circles show the stars observed in the VLT/FLAMES low resolution \ion{Ca}{ii} triplet survey~\citep{Battaglia07,Starkenburg10}. The~(red) solid dots mark the RGB stars that also have high resolution abundance measurements~\citep[Hill et al., in prep, see][]{Tolstoy09} and the~(green) dashed ellipse is the tidal radius of Sculptor, as given by~\citet{Irwin95}. \label{Sclcov}} 
\end{figure}
\\
The presence of these gradients suggests that Sculptor has experienced an initial, spatially extended episode of metal-poor star formation at all radii, and stars of higher metallicity were subsequently formed more towards the centre. This picture explains the properties of the CMD, such as the different spatial distributions traced by the HB morphology and Red Giant Branch~(RGB) metallicity gradient. \\
Sculptor is a galaxy that can be used as a benchmark for a simple galaxy with an extended episode of star formation in the early Universe. However, in order to quantify this picture and obtain accurate timescales for different stellar populations at different radii the detailed SFH needs to be determined over a large area of the galaxy. \\
The Sculptor dSph has been modelled several times in simulations using different techniques~\citep[e.g.,][]{Salvadori08, Revaz09, Lanfranchi10, Kirby11,Revaz11}. The cosmological semi-analitycal model of~\citet{Salvadori08} follows simultaneously the evolution of the Milky Way and its dwarf galaxy satellites, and reproduces the observed MDF and total mass content of the Sculptor dSph. \citet{Lanfranchi10} use chemical evolution modelling to reproduce the observed MDF of a variety of dwarf spheroidal galaxies, including Sculptor~\citep[see also][]{Lanfranchi04}. \citet{Kirby11} use a chemical evolution model to match the observed MDF and alpha-element distributions from a large sample of spectroscopic observations in Local Group dwarf spheroidal galaxies. \citet{Revaz11} use a chemo-dynamical Smoothed-particle Hydrodynamics~(SPH) code to model the properties of Local Group dwarf spheroidal galaxies. Their model for the Sculptor dSph correctly matches the observed MDF and shows an extended SFH. Furthermore, the narrow [Mg/Fe] distribution in their Sculptor-like model matches well with observations from HR spectroscopy. \\
In this paper the SFH of a large area of  the Sculptor dSph will be determined using CMD synthesis methods~\citep[e.g.,][]{Tosi91,Tolstoy96,Gallart962,Dolphin97,Aparicio97} to interpret the deep wide-field photometry presented in~\citet{deBoer2011A}. The available spectroscopic information from \ion{Ca}{ii} triplet and HR spectroscopy will be directly used together with the photometry to provide additional constraints on the SFH. The spatial coverage of the photometric data~(covering~$\approx$80\% of the tidal radius of Sculptor) allows us to determine the SFH at different radii and quantify the observed radial age and metallicity gradients. \\
Using the detailed SFH we also determine the probability distribution function for age for stars on the upper RGB, giving age estimates for individual stars. By linking these ages to the observed spectroscopic abundances we directly obtain, for the first time, the timescale of enrichment from different types of Supernovae~(SNe). \\
The paper is structured as follows: in section~\ref{data} we present our photometric and spectroscopic observations. In section~\ref{method} we describe our method of obtaining the SFH using the synthetic CMD method, adapted to include the MDF information. The detailed SFH analysis of the Sculptor dSph is given in section~\ref{results}. The details of the age determination of individual stars are given in Section~\ref{indivages}. Finally, the conclusions that can be drawn from the SFH are discussed in section~\ref{conclusions}.

\section{Data}
\label{data}
Deep optical images of the Sculptor dSph in the B, V and I bands were obtained using the CTIO 4-m MOSAIC II camera. The reduction and accurate calibration of this dataset is described in detail in a preceding paper~\citep{deBoer2011A}. The total coverage of the nine photometric fields observed is shown in Figure~\ref{Sclcov} as the solid black square. The spatial coverage of the B band photometry is complete for radii r$_{ell}$$\le$0.5 degrees, while the V and I bands are complete for r$_{ell}$$\le$1 degree. \\ 
By stacking together several images for each pointing the deepest photometry possible was obtained. Short exposures were also obtained, to be able to accurately photometer the bright stars that are saturated in the deep images. In order to ensure accurate photometric calibration, images were obtained with the 0.9m CTIO telescope under photometric conditions. Observations were also made of Landolt standard fields~\citep{Landolt07, Landolt92}. Photometry was carried out on these images using DoPHOT~\citep{Schechter93}. The different fields were placed on the same photometric scale and combined in order to create a single, carefully calibrated photometric catalog, as described in~\citet{deBoer2011A}. 
\begin{figure}[!ht]
\centering
\includegraphics[angle=0, width=0.49\textwidth]{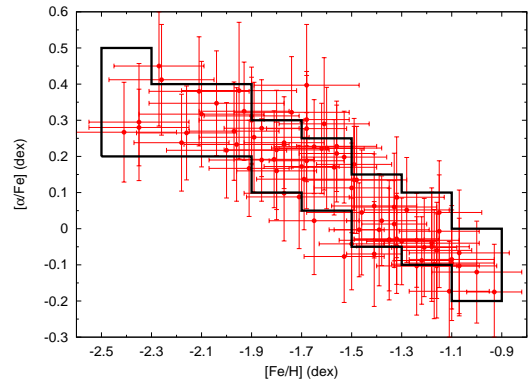}
\caption{Spectroscopic [Fe/H] and [$\alpha$/Fe] measurements with their errors obtained from HR observations~(Hill et al., in prep) in the central 25$^{\prime}$ diameter of the Sculptor dSph. The solid~(black) line shows the range of possible values of [$\alpha$/Fe] assumed for each bin of [Fe/H] in the SFH determination. \label{fealprel}} 
\end{figure}
\\
In addition, \ion{Ca}{ii} triplet spectroscopy is available for~$\approx$630 individual RGB stars that are likely members of Sculptor, from medium resolution~(R$\sim$6500) VLT/FLAMES observations~\citep{Battaglia07,Battaglia082,Starkenburg10}. These observations measure [Fe/H] for a large sample of stars for which we also have photometry, out to a radius of 1.7 degrees from the centre of the Sculptor dSph~(see Figure~\ref{Sclcov}).  \\
Furthermore, HR spectroscopy~(VLT/FLAMES) is available for 89 individual RGB stars within the central part of the Sculptor dSph, for r$_{ell}$$\le$0.2 degrees~\citep[Hill et al., in prep, see][]{Tolstoy09}. These observations provide [Fe/H] as well as [$\alpha$/Fe] measurements. For [$\alpha$/Fe] we assume [$\alpha$/Fe] =([Mg/Fe] +[Ca/Fe] +[Ti/Fe])/3. The HR spectroscopy includes a range in metallicity from $-$2.5$<$[Fe/H]$<$$-$1.0 dex, with alpha-element abundances showing a clear correlation with [Fe/H] for the range $-$0.3$<$[$\alpha$/Fe]$<$0.5 dex~(See Figure~\ref{fealprel}). 

\section{Method}
\label{method}
CMDs contain the signatures of numerous evolutionary parameters, such as age, chemical abundance, initial mass function, etc. The typical method of determining a SFH relies on comparing observed and synthetic CMDs of the individual stars that can be resolved. There have been many schemes proposed to quantify the SFH extracted from the CMD~\citep[e.g.,][]{Tosi91,Tolstoy96,Gallart962,Dolphin97,Aparicio97,Dolphin02}. \\
We have created our own routine, Talos, which uses an approach that compares observed CMDs with a grid of synthetic CMDs through the use of Hess diagrams (density plots of stars in the CMD) to determine the SFH~\citep[similar to][]{Dolphin97,Dolphin02}. The synthetic CMDs include observational effects in a statistical manner, which provides the most realistic way to include them. \\
We assume that a SFH can be built up from a linear combination of simple stellar populations. The advantage of this approach is that instead of synthesising a large number of artificial CMDs, each with their own complex Star Formation Rate~(SFR(t)), the simple populations only need to be generated once. Then, the combination of simple stellar populations has to be found, which best represents the observed CMD. \\
Using the standard technique a number of different CMDs~(e.g., V,B-V and I,V-I) can be independently used to obtain SFHs. A CMD is inherently two dimensional~(colour and magnitude), whereas when photometric information is available in more than two filters, we can use 3D information to more precisely constrain the SFH. \\
To fully utilize all the available photometric information to constrain the SFH, Talos fits the measurements in all the available passbands at once. In this way all the advantages of the different CMDs~(such as the more precise photometry in the I,V-I CMD and the larger colour range of the V,B-I CMD) are incorporated into a single, more accurate SFH. \\
In addition to the photometric data we can also add spectroscopic observations of a large number of individual RGB stars. To take into account this extra information Talos also fits the observed MDF at the same time as the photometry, which allows us to put well motivated constraints on the metallicity range of the different stellar populations. \\
The SFH determination technique consists of the following steps:
\begin{enumerate}
\item Construct synthetic CMD and MDF models
\item Find the combination of models that best match the data.
\item Determine the resolution and uncertainties of the final SFH.
\end{enumerate}
These steps will be explained in more detail in the following sections.

\subsection{Constructing synthetic CMD models}
\label{CMDmodels}
The synthetic stellar evolution library adopted in Talos is the Darthmouth Stellar Evolution Database~\citep{DartmouthI}. The database contains isochrones distributed with a specific grid of age, metallicity and $\alpha$-element abundance, among which are also those with [$\alpha$/Fe]$<$0. These are of particular importance for our analysis, since Sculptor contains a substantial number of stars~(33\% of the total) with low $\alpha$-element abundances~(See Figure~\ref{fealprel}). \\ 
In order to allow greater flexibility in choosing the population domain and parameter resolution to use when modelling the SFH, we interpolate between the provided isochrones to produce a finer grid. A routine is provided by the Dartmouth group, to interpolate isochrones in [Fe/H]. Interpolation in age and [$\alpha$/Fe] is done linearly, given that the grid of age points is sufficiently fine that there will only be small changes between isochrones of different age. \\
The isochrone library does not model the HB, Asymptotic Giant Branch~(AGB) or Blue Straggler Star~(BSS) phases, which means we cannot use these evolutionary features to constrain the SFH. The AGB is especially important, since it merges with the RGB in the CMD. The largest uncertainties in our final model may come from misidentifying the AGB using RGB models or the BSS as a young population. However, the Dartmouth isochrones have been shown to give a good simultaneous fit to all other evolutionary features within a CMD, justifying their use~\citep[e.g.,][]{Glatt2008a,Glatt2008b}. \\
Synthetic CMDs are generated by drawing stars from the isochrones according to an initial mass function~(IMF) and the mass range within each isochrone. The IMF used is the Kroupa IMF~\citep{Kroupa01}. An initial SFR for a stellar population~(within a default mass range of 0.1$-$120~M$_{\odot}$) is assumed~(resulting in several times more synthetic stars than in the observed CMD), which ensures that enough stars are generated in each part of the synthetic CMD, consistent with the time spent in each evolutionary phase. These synthetic stars are then placed at the distance of the Sculptor dSph, assuming a reddening of~E(B-V)=0.018. \\
This gives us ideal CMDs for each stellar population that we consider in the SFH analysis. To create a synthetic CMD that can be compared directly  to the observed photometry we need to include observational effects in order to simulate our observational limitations. 

\subsubsection{Adding observational effects}
The crucial aspect in comparing observed and synthetic CMDs is to determine the observational biases, such as photometric errors, incompleteness, etc. \\
This is done by carrying out a large number of simulations, in which a number of artificial stars~(with known brightness) are placed on the observed images. These images are then re-reduced in exactly the same way as the original images, after which the artificial stars are recovered from the photometry. In this way we obtain a look-up table, which can be used to accurately model the effects of observational conditions~\citep[e.g.,][]{Stetson88,Gallart961}. The lookup-table is used in Talos to assign an individual artificial star~(with similar colours and magnitudes) to each star in an ideal synthetic CMD and considering the manner in which this star is recovered to be representative of the effect of the observational biases.
\begin{figure}[!htb]
\centering
\includegraphics[angle=0, width=0.495\textwidth]{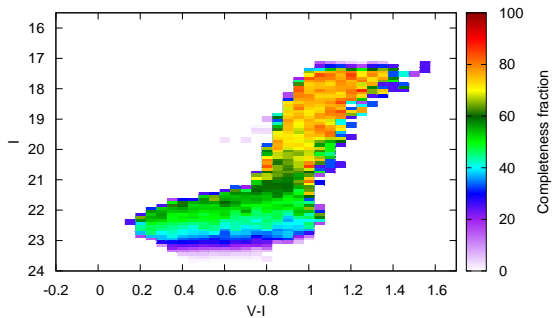}
\caption{The recovered I,V-I CMD of the artificial stars put into the observed images of Sculptor. The colours indicate the completeness fraction of artificial stars in each bin, with a scale on the right hand side of the plot. \label{SclcompHessVI}} 
\end{figure}
\\
Using individual artificial stars has the disadvantage that a huge number of stars need to be generated in order to correctly sample the effects in each CMD bin. This is because using the same artificial star in assigning offsets to several synthetic stars creates artificial clustering in the CMD. However, this approach is the only way to properly take into account the colour-dependence of the completeness level and the asymmetry of the offsets applied to model stars at faint magnitudes~\citep[e.g.][]{Gallart961}.  
\begin{figure}[!ht]
\centering
\includegraphics[angle=0, width=0.45\textwidth]{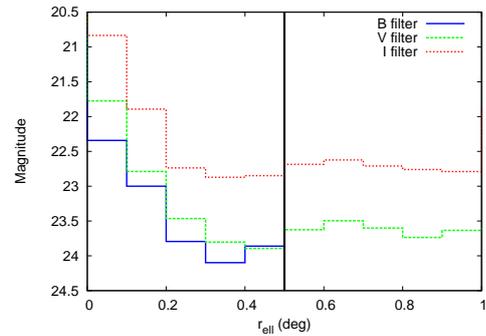}
\caption{The magnitude of the 50\% completeness level with increasing elliptical radius~(r$_{ell}$) from the centre of the Sculptor dSph, in B~(solid blue line), V~(green dashed line) and I~(red dotted line) filters. The vertical~(black) line indicates the elliptical radius up to which observations in the B filter are available. \label{radialcomp}} 
\end{figure}
\\ 
The set of artificial stars was generated with parameters encompassing the range 5$<$Age$<$15 Gyr, $-$2.5$<$[Fe/H]$<$$-$0.80 dex, $-$0.2$<$[$\alpha$/Fe]$<$0.40 dex. The stars were distributed randomly across the nine MOSAIC pointings in Sculptor, in 3 different filters. In each image no more than 5\% of the total observed stars were ever injected as artificial stars at one time, so as not to change the crowding properties in the image. For each observed pointing 200$-$400 images containing artificial stars were created, in order to obtain a sufficient total number of artificial stars. This resulted in nearly 7000 images containing a total number of 3.5 million artificial stars spread across the full area of the Sculptor dSph. Each image was re-reduced using the same techniques and calibrations as the original image, after which the output catalog was matched to the input catalog~(while taking care not to match artificial stars to observed real stars), to produce a lookup-table containing the recovered artificial stars~(and the recovered photometric offsets) in 3 filters. \\
Figure~\ref{SclcompHessVI} shows the result of the artificial star tests in the I,V-I CMD of Sculptor. The completeness level accounts for only those objects that DoPHOT unambiguously recognises as stars. DoPHOT actually detects more objects, but the faintest objects do not have high enough signal-to-noise to be confidently distinguished as stars as opposed to unresolved galaxies. Only the unambiguously detected stars are considered in the completeness level, since those are observed with sufficient accuracy to be useful in our analysis. Figure~\ref{SclcompHessVI} shows the dependence of the completeness level on both colour and magnitude. The dependence on colour in particular has to be properly taken into account when generating realistic synthetic CMDs. \\
The need to carry out artificial star tests over the entire observed field of view is highlighted in Figure~\ref{radialcomp}. Here we show the variation of the 50\% completeness level with distance from the centre of Sculptor, in the three different filters. The completeness level in the B filter is zero for r$_{ell}$$\ge$0.5 degrees, because of the lack of B band observations in the outer parts of Sculptor. Figure~\ref{radialcomp} also shows that the centre of Sculptor is less complete at a fixed magnitude limit than the outer regions. This is due to the increased crowding in the central region, which means it is harder to determine the shape of the PSF of stars in the centre, causing fewer stars to be unambiguously detected at a given brightness level, and placing the 50\% completeness at brighter magnitude levels. \\
However, the CMDs at different radii from the centre~\citep[see Figure~5 from][]{deBoer2011A} show that the photometry in the central part of Sculptor actually goes deeper than in the outskirts~(due to longer exposure times), with a well defined shape of the MSTOs. Additionally, using the artificial star test results to simulate observational conditions makes sure that this changing completeness level is properly taken into account when determining the SFH. \\
Artificial star tests have only been carried out on the long, stacked exposures. For those stars saturated in the long exposures, observational effects are modelled by giving them random offsets within the photometric error distribution of stars with similar instrumental magnitude and colour. This is justified because the recovery fraction for the brightest stars is very close to 100\%, and the offsets induced by crowding  are insignificant and signal-to-noise is not an issue. \\
The process of including observational errors in synthetic CMDs in a statistical manner means that they can be directly compared to the observed CMDs, to obtain the best matching SFH. 

\subsection{Constructing MDF models}
\label{MDFmodels}
In order to include the spectroscopic MDF at the same time as the photometry in Talos we use the equivalent of a Hess diagram for [Fe/H]. This is done by generating synthetic CMDs of different stellar populations and binning the stars in metallicity. In this way it is possible to create synthetic MDF models for stellar populations of different [Fe/H], age and [$\alpha$/Fe]. \\
The spectroscopic observations only come from a fraction of the RGB, which are in turn a small fraction of stars present in the CMD. To make sure that the synthetic CMDs fully sample the RGB, synthetic MDF populations are generated with an artificially increased initial SFR, which is later corrected in the MDF models. To correctly reproduce observational limits we construct the synthetic MDFs using only stars within a similar magnitude range to those of the the observed spectroscopic sample. From the number of stars observed both photometrically and spectroscopically in the same magnitude range we then calculate the spectroscopic completeness fraction on the RGB. This fraction is used to scale down the synthetic MDF, and thus match the spectroscopic completeness. \\
The observed spectroscopic uncertainties on the metallicites are simulated by considering each individual synthetic star to have a Gaussian profile with a width determined by the average observational uncertainty on [Fe/H]. The metallicities of these individual stars are then combined to form a synthetic MDF which takes into account observed uncertainties, and can be directly compared to the observed MDF.  

\subsection{Determining the SFH}
In Section~\ref{CMDmodels} we described how Talos can be used to create accurate synthetic CMDs, which can be compared directly to the observed CMDs. We then described in Section~\ref{MDFmodels} how model MDFs are created in Talos, which add additional information that can be used to restrict the SFH. Here we now put everything together to determine the SFH of a galaxy. \\
The technique used to determine the best SFH minimises the difference between observed and synthetic CMDs through Hess diagrams to produce the closest match to the observed data, as described in~\citet{Dolphin02}. Photometric Hess diagrams are constructed by binning CMDs in magnitude and colour space~(or three magnitude space for Hess diagrams using B,V and I filters) and counting the number of stars in each bin. The synthetic Hess diagrams are scaled to contain a number of stars equal to the observed Hess diagrams, to make sure that no model is given preference during the SFH determination based on the total number of stars it contains. \\
In Talos, the synthetic MDFs are used in the fitting procedure in a similar manner as the photometric Hess diagrams. The spectroscopic MDF is treated as an extra observed dimension, for which the difference between model and observations also needs to be taken into account. A different weight is applied to the photometric and spectroscopic components, in order to enhance the importance of the spectroscopic information, which contributes less to the overall goodness of fit than the photometry, but significantly restricts the possible solutions. \\
Since the observed data follows a Poisson distribution, the goodness of fit is expressed as a difference parameter between models and observations~(a Poisson equivalent of $\chi^{2}$), given by~\citet{Dolphin02}:
 \begin{displaymath}
\chi_{Poisson}^{2} = 2\sum_{i}^{}m_{i} - n_{i} + n_{i} ln\frac{n_{i}}{m_{i}}
 \end{displaymath}
\\
In which $m_{i}$ is the total number of stars in a synthetic Hess bin~ ($m_{i}=\sum_{j}^{}$SFR$_{j} \times$CMD$_{i,j}$ in the photometric part, and $m_{i}=\sum_{j}^{}$SFR$_{j} \times$MDF$_{i,j}$ in the spectroscopic part) and $n_{i}$ the total number of stars in an observed Hess bin. By minimising the difference parameter we determine the SFH~(given by SFR$_{j}$) that best matches both the observed CMD as well as the observed spectroscopic MDF.

\subsubsection{SFH uncertainties}
\label{uncertainties}
Equally as important as determining the best matching SFH is determining the uncertainties on this choice. The two main sources of statistical uncertainties on the determination of the SFH are related to data and parameter sampling of the SFH solution~\citep{Aparicio09}. 
\begin{figure}[!htb]
\centering
\includegraphics[angle=0, width=0.45\textwidth]{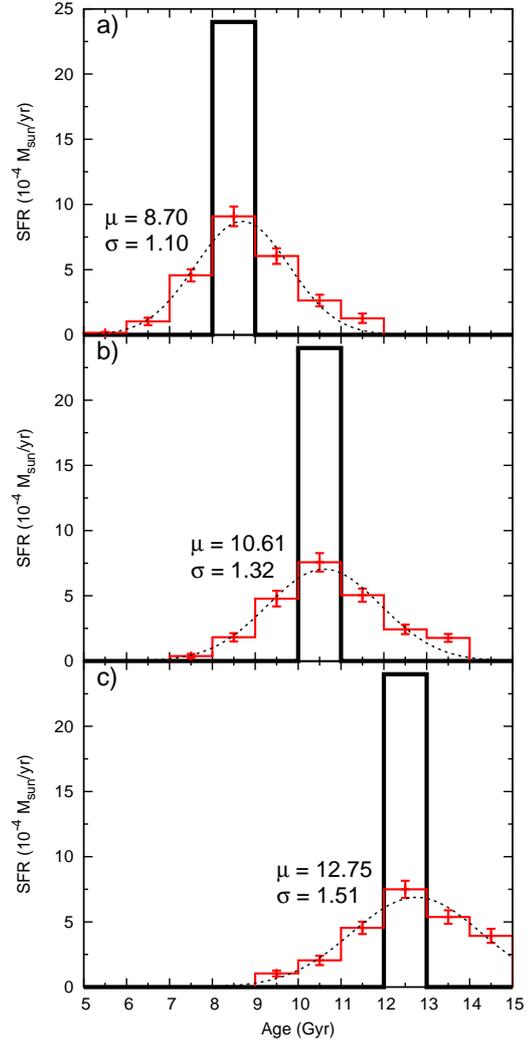}
\caption{The input and recovered SFHs of a series of short~(10 Myr) bursts of star formation at different input age. The black, solid histogram shows the input SFH, given the binning adopted. The red histograms show the recovered SFH, along with the fit of a Gaussian distribution as the dashed line. The mean~($\mu$) and variance~($\sigma$) of the fitted Gaussian distribution are also listed. \label{SFHresolution}} 
\end{figure}
\\
Data sampling is related to how representative a model is of the underlying population. In the context of CMD analysis, constructing the same model twice, with different random samplings of the same isochrones will result in two slightly different synthetic CMDs. In order to quantify how this effect may impact the inferred SFH without synthesizing an entire new set of synthetic CMDs we generate a number of different representations of the observed CMD. Synthetic CMDs are generated from the SFH obtained in the first run-through of the code and randomly swapping a certain fraction~(by default 20\%) in the observed CMD. In this way we obtain a new ``observed" CMD which is representative of a CMD with a different sampling of the underlying population. By generating a number of these CMDs and using them with the original models to find the SFH we can quantify the errors on the SFH caused by data sampling~\citep{Aparicio09}. \\
The other major source of error is parameter sampling. This means that the choice of CMD gridding, age bins and metallicity bins will affect the final SFH. In order to quantify the effect of this gridding on the recovered SFH, different bin sizes and distributions are used and then compared. In the parameter space of age and metallicity three different shifts are applied: a shift of half a bin size in age and in metallicity and also a shift in both age and metallicity simultaneously. For the photometric binning two different bin sizes are used, with the different shifts applied to each. For all these different griddings the SFH is determined, after which the results are returned to a common grid. \\
The average of all the different solutions is adopted as the final SFH, with errorbars determined as the standard deviation of the distribution of solutions. This technique has been shown to adequately take into account the major uncertainties on the recovered SFH, and give realistic errorbars on the final SFH~\citep{Aparicio09}. 
\begin{figure*}[!ht]
\centering
\includegraphics[angle=0, width=0.45\textwidth]{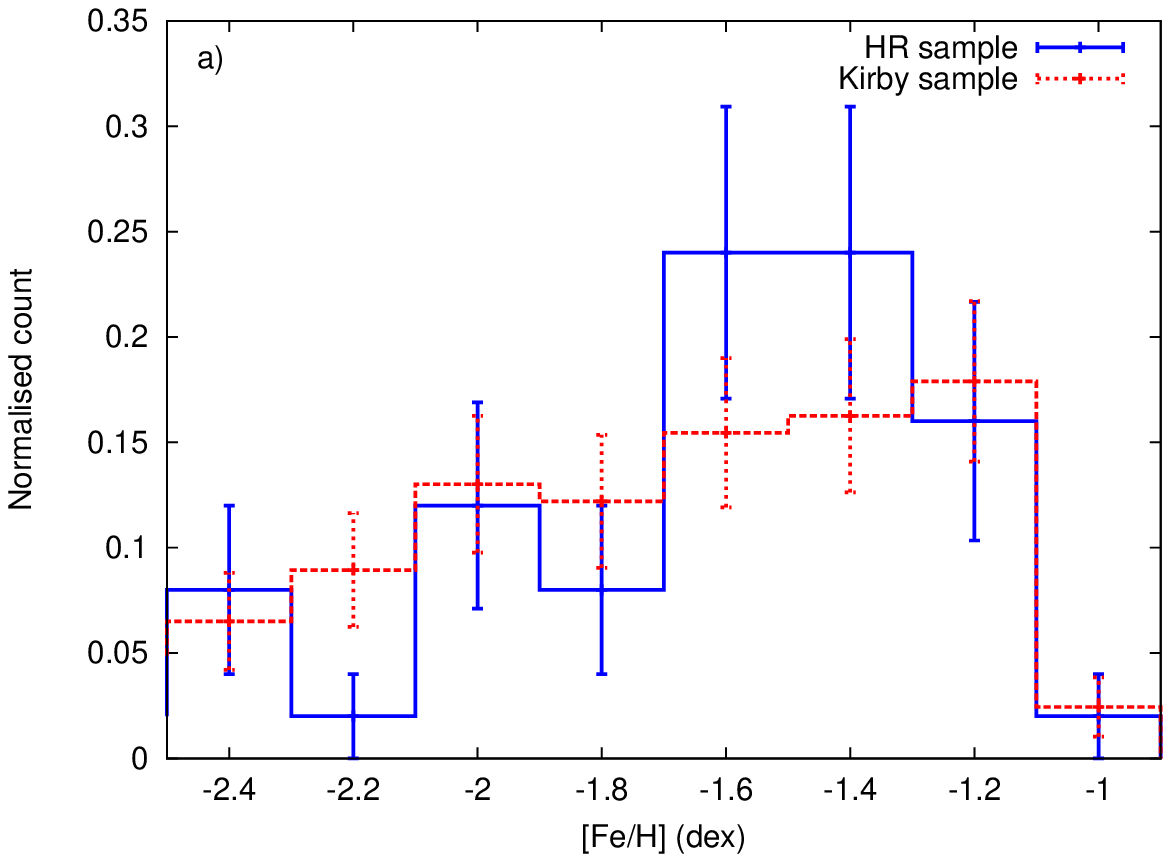}
\includegraphics[angle=0, width=0.45\textwidth]{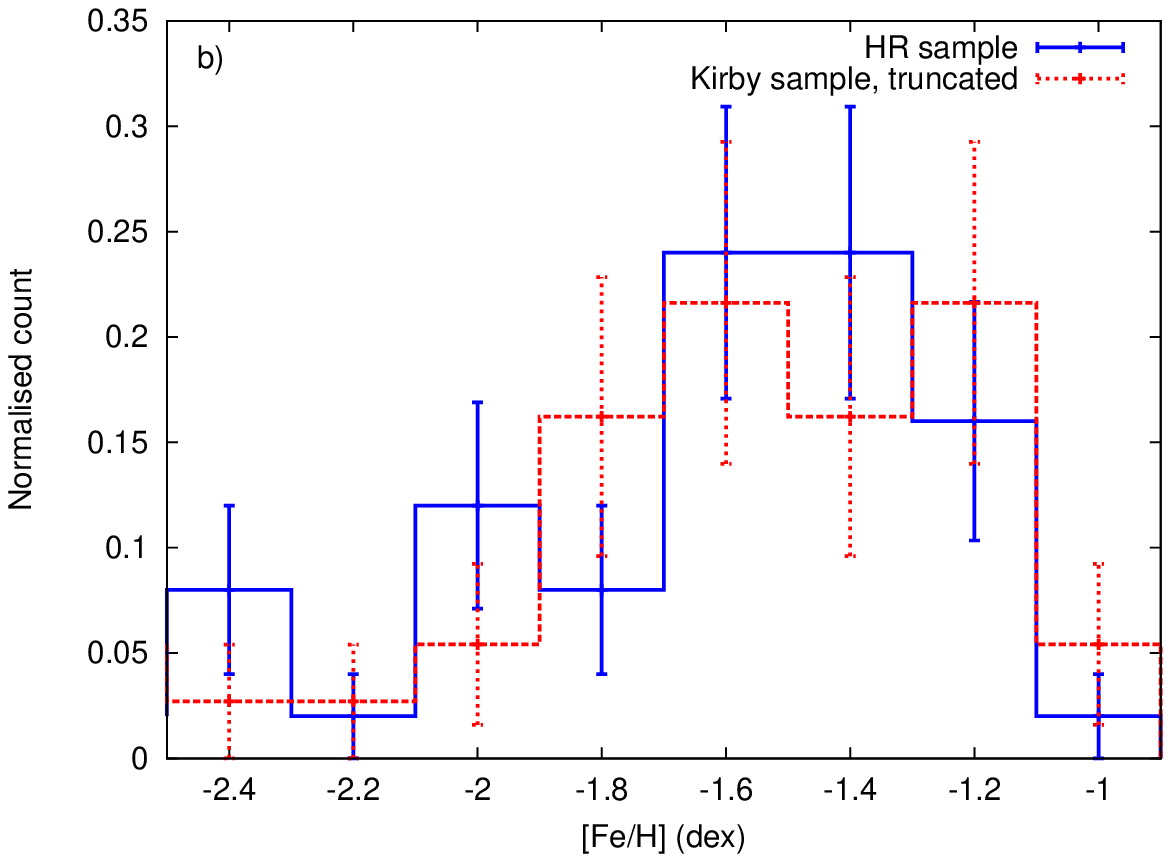}
\caption{The MDF of RGB stars in the Sculptor dSph from~Hill et al., in prep~(HR sample, blue solid histogram) and~\citet{Kirby10}~(red dashed histogram). Panel~\textbf{a)} shows the complete samples for both MDFs~(HR, going down to V$\approx$18 and Kirby et al. going down to V$\approx$20), while panel~\textbf{b)} shows the Kirby MDF truncated at V$\approx$18 along with the full HR sample. Poissonian errorbars are also shown. \label{MDFcomparison}} 
\end{figure*}

\subsection{General properties of Talos}
\label{Scltests}
In order to verify the basic operation of Talos, a number of tests have been performed, and the results are shown in Appendix~\ref{tests}. We checked our ability to correctly recover the age and metallicity of a simple synthetic stellar population. We also showed that we can accurately recover a synthetic SFH of a continuous period of star formation from the earliest times until now. Furthermore, we tested Talos on observations of the globular cluster NGC1904. The metallicity we determined is consistent with that obtained from spectroscopic observations, showing that Talos can accurately recover the age and metallicity of an observed simple stellar population. The main effect that makes the output solution different from the input is the depth of the photometric data, and the set of parameter griddings adopted to compute the final SFH. These result in limits to the age resolution of our final SFH. \\
In order to understand the limitations of the final SFH determined by Talos, it is important to obtain an accurate estimate of the age resolution of the solution. This is determined by recovering the SFH of a set of synthetic populations at different input ages, as described by~\citep[e.g.,][]{Hidalgo11}. \\
Three synthetic populations were generated at the distance of the Sculptor dSph, with single short bursts of star formation~(with a duration of 10 Myr) at an age of 8.5, 10.5 and 12.5 Gyr. The metallicity of the bursts was distributed to match the observed MDF of the Sculptor dSph, as were the observational errors. \\
The recovered and input SFH for these bursts are shown in Figure~\ref{SFHresolution}. The recovered SFH is well fit by a Gaussian distribution, which shows that the mean of the central peak~($\mu$) is recovered at the correct age, with a minor shift of~$\approx$0.1$-$0.2 Gyr. The recovered SFH shows that typically~$\approx$40\% of the total input SFR is contained within the central bin, at all ages. The star formation is spread out over the Gaussian distribution, leading to uncertainties in the recovered SFH. However, in the case of constant star formation~(see Section~\ref{synthetictests}) the recovered star formation rates show a more accurate recovery of the input SFH. \\
From the Gaussian fits to the three bursts in Figure~\ref{SFHresolution} we can derive the variance~$\sigma$, which determines the resolution with which the burst is recovered. The three bursts are recovered with an age resolution of~$\approx$1.5 Gyr at an age of 12.5 Gyr,~$\approx$1.3 Gyr at an age of 10.5 Gyr and~$\approx$1.1 Gyr at an age of 8.5 Gyr, which is consistent with a value of 12\% of the adopted age.
\begin{figure}[!ht]
\centering
\includegraphics[angle=0, width=0.475\textwidth]{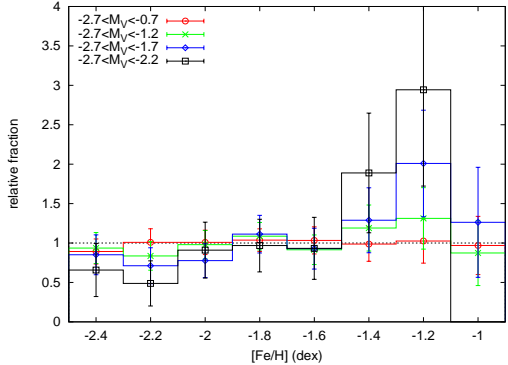}
\caption{The relative fractions of different metallicities present on the RGB at different photometric depths, with respect to an MDF going down to M$_{V}$=$-$3.2~(V=19.5) on the RGB. The relative fraction indicates by which factor a population is incorrectly sampled in the CMD. For comparison the level of the HB is V$_{HB}$=20.13~(M$_{V}$=0.43). \label{lumfuncfrac}} 
\end{figure}

\subsection{Simulating the Sculptor dSph}
\label{Sclspecifics}
To use Talos to determine the SFH of the Sculptor dSph the general method described in the preceding sections needs to be adapted to the specifics of this galaxy. 

\subsubsection{General setup}
The observed CMD of Sculptor contains both stars that belong to Sculptor as well as Milky Way foreground and unresolved background galaxies. To avoid these objects influencing the SFH, the observed CMD is ``cleaned" using the colour-colour diagram presented in~\citet{deBoer2011A}. Foreground stars with colours that coincide with those of Sculptor are assumed not to significantly alter the total number of stars in different evolutionary features, as determined by using the Besan\c{c}on models to predict the number of Milky Way stars in the CMD~\citep{Robin03}. \\
Additionally, the observed CMDs~\citep[shown in Figure~5 from][]{deBoer2011A} show that Sculptor contains a significant number of BSS stars, which we do not attempt to include in the SFH determination as a younger population, since they are likely to be genuine BSS stars~\citep{Mapelli09}. To avoid the BSS stars influencing the SFH, we generate a mock BSS population which is used as a fixed background in the SFH determination. The mock BSS stars are constructed by using only the main sequence part of populations spanning a range in metallicity of $-$2.5$<$[Fe/H]$<$$-$1.0, with ages between 3$-$4 Gyr. This background Hess diagram is subsequently scaled so the number of synthetic stars in the BSS region matches the observed number. 
\begin{figure*}[!ht]
\centering
\includegraphics[angle=0, width=0.49\textwidth]{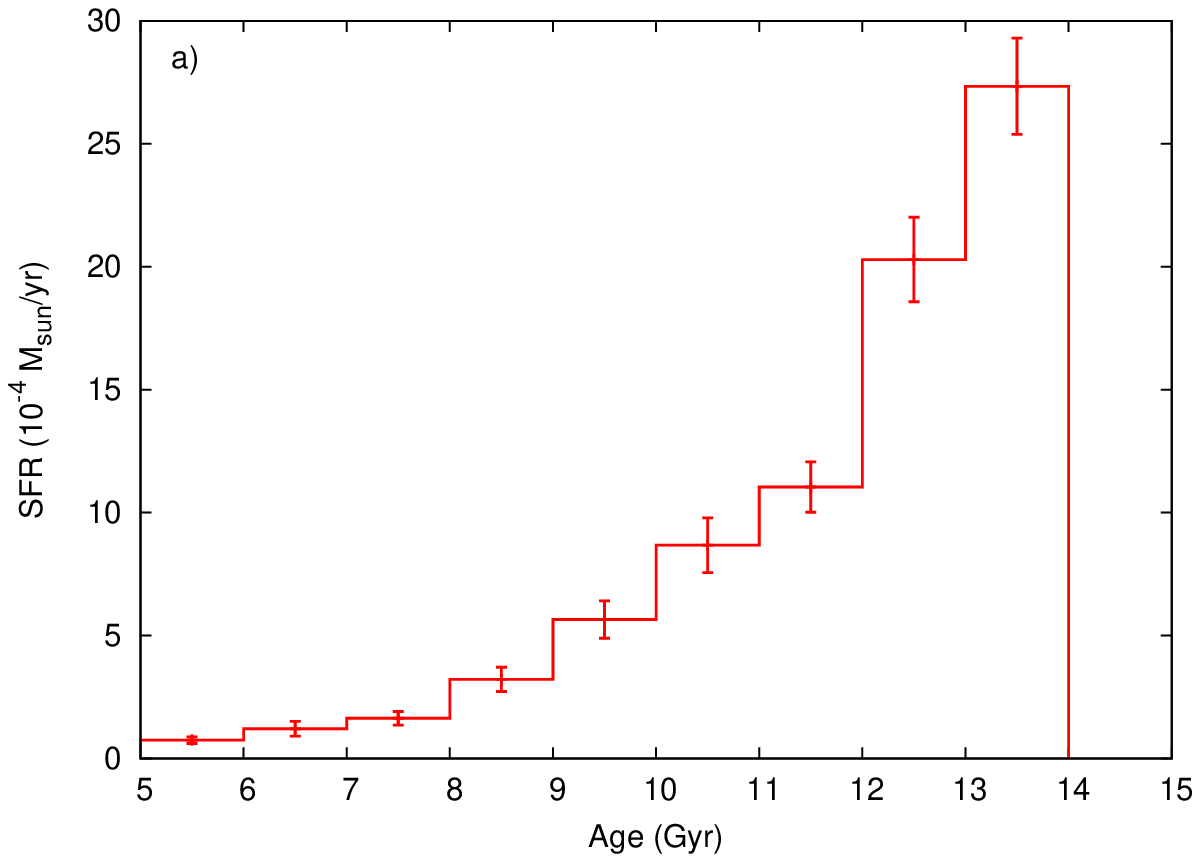}
\includegraphics[angle=0, width=0.49\textwidth]{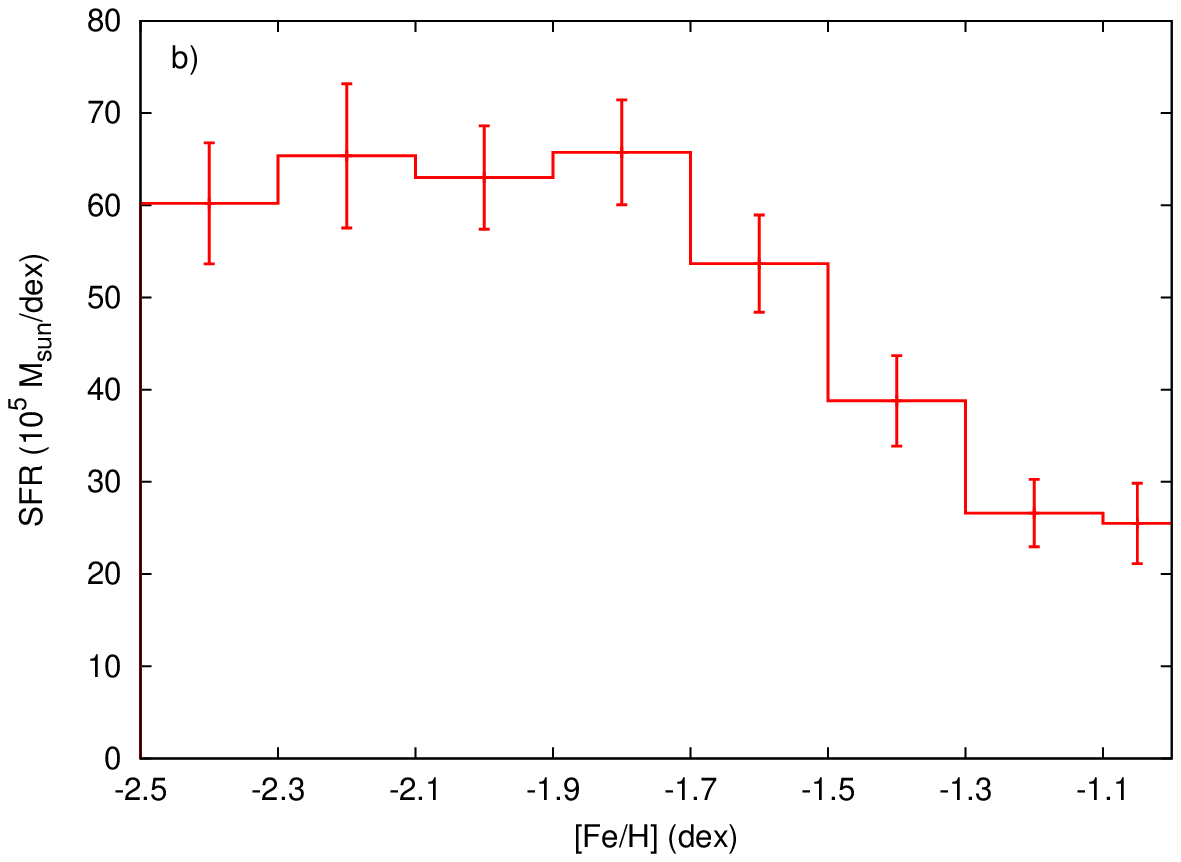}
\caption{The final~(\textbf{a}) SFH and~(\textbf{b}) Chemical Evolution History of Sculptor, out to a radius of r$_{ell}$=1 degree, obtained by fitting all available photometric and spectroscopic information simultaneously. \label{overallSFH}} 
\end{figure*}
\\
The SFH is determined using the CMD without including the RGB region. The number of stars on the RGB is low compared to that in the MSTO region, and therefore does not contribute strongly to the determination of the SFH. Furthermore, the RGB feature has been shown to substantially vary from one set to the other~\citep[e.g.,][]{Gallart05}. Conversely, there is good agreement between the theoretical MSTO phase at old ages~($\ge$10 Gyr) in different isochrone sets. Instead, information from the RGB phase is taken into account through the inclusion of the MDF fitting. \\
Determining the SFH without including the spectroscopic MDF shows similar overall trends as obtained when including the MDF. However, the addition of the MDF has a result of putting better constraints on less populated region of the CMD, such as the metal-rich stars, which would otherwise be fit using a lower metallicity and higher age, due to the mismatch between the RGB and MSTO in the isochrones. \\
A single, well defined distance is assumed for the Sculptor dSph in the SFH determination, as obtained by a number of reliable distance indicators~\citep{Pietrzynski08}. The effect of adopting a different distance is shifting the main peak of star formation to younger ages for a larger distance and older ages for a smaller distance. However, the internal distribution of the presented SFHs does not change significantly when adopting a different distance, given the small observed uncertainties on the distance. Furthermore, the good agreement between different distance determination methods gives confidence to the single distance adopted in this work.

\subsubsection{Taking into account the spectroscopic information}
The spectroscopic information which is used to provide extra constraints on the SFH comes from two different sources. For r$_{ell}$$\le$0.2 degrees the metallicity information comes from HR spectroscopy on the upper RGB, while in the outer parts of Sculptor~(0.2$\le$r$_{ell}$$\le$1.7 degrees) it comes only from \ion{Ca}{ii} triplet spectroscopy which also includes fainter stars. \ion{Ca}{ii} triplet spectroscopy is also available in the central region, which shows that metallicities from both samples are placed on the same scale, and consistent with each other~\citep{Battaglia082}. However, because the spectroscopic sample in the central region encompasses only the brightest $\approx$1mag of the RGB, an intrinsic bias may be present in the observed central MDF due to the limited sampling of the stellar populations on the upper RGB.  \\ 
To check for the presence of such a bias in our HR spectroscopic sample, a comparison is made with a different spectroscopic sample taken in the same region, going down to fainter magnitudes~\citep{Kirby10}. Figure~\ref{MDFcomparison}a shows a comparison between the normalised MDFs of both samples. The MDF of the deeper sample clearly shows a larger fraction of metal-poor stars than the brighter sample. Figure~\ref{MDFcomparison}b shows that the MDF of both samples look the same, when the sample of RGB stars is truncated at the same magnitude. Figure~\ref{MDFcomparison} shows that an MDF determined from only the upper RGB stars in Sculptor results in a lower relative fraction of metal-poor stars. This effect is due to the luminosity function bias on the upper RGB, causing an incomplete sampling of the metal-poor components for the brightest RGB stars. \\ 
In order to investigate the effects of the bias in more detail we consider the sampling of the MDF at different photometric depths. Using a synthetic CMD of a Sculptor-like galaxy for which the metallicity is known for each star, we construct MDFs from RGB stars going down to different photometric depths. By comparing to the input parameters of the synthetic CMD we find that going down to M$_{V}$=$-$3.2~(V=19.5) fully samples the overall MDF. MDFs of different depths are compared to the overall MDF to produce the correct relative fraction of stars sampled at each metallicity. Figure~\ref{lumfuncfrac} shows the relative fraction of stars sampled on the RGB at different metallicities in CMDs of different photometric depth. The effect of depth on the RGB is clearly seen, with shallower MDFs under-sampling populations with [Fe/H]$<$$-$1.9 dex and over-sampling more metal-rich components. A similar effect is also seen when applying the same approach to the observed, deep \ion{Ca}{ii} triplet data, with a more severe under-sampling of the populations with [Fe/H]$<$$-$1.9 dex compared to the synthetic CMD results. \\
It is important not to neglect the metal-poor component in Sculptor, which is well sampled by the MSTO region, but not on the RGB. Figure~\ref{lumfuncfrac} shows that no bias is present in the MDF for [Fe/H]$\ge$$-$1.9 dex. Therefore, we adopt this as a metallicity cut-off, below which the SFH is only constrained using photometry. In the outer parts of Sculptor~(r$_{ell}$$\ge$0.2 degrees) no metallicity cut-off is adopted, since the \ion{Ca}{ii} triplet spectroscopy extends to faint enough magnitudes to avoid this bias. 
\begin{figure*}[!ht]
\centering
\includegraphics[angle=0, width=0.45\textwidth]{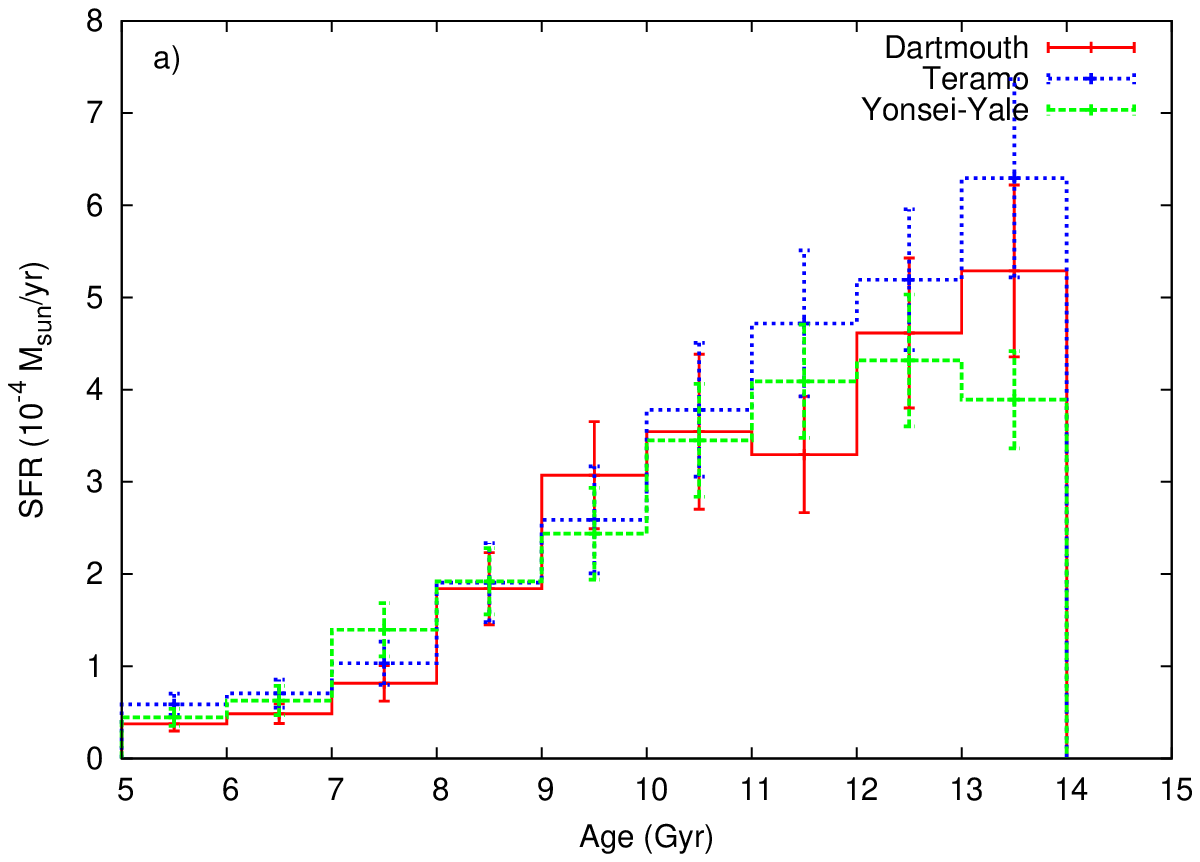}
\includegraphics[angle=0, width=0.45\textwidth]{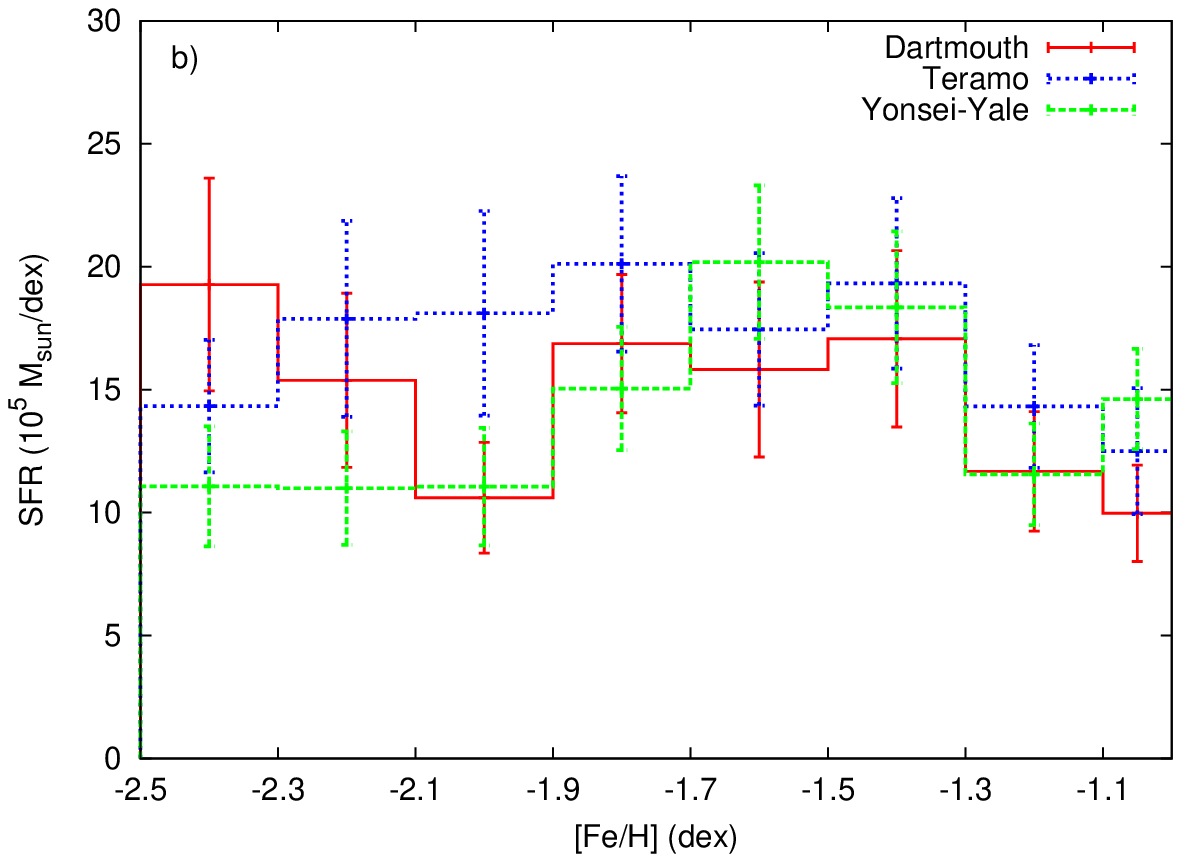}
\caption{The~(\textbf{a}) SFH and~(\textbf{b}) CEH of the centre of Sculptor, determined using three different isochrone sets. The obtained solution is shown for the Darthmouth Stellar Evolution Database as the~(red) solid line, for the BaSTI/Teramo isochrones as the~(blue) dotted line and for the Yonsei-Yale isochrones as the~(green) dashed line. \label{isotest}} 
\end{figure*}

\subsubsection{SFH parameter space}
To set the limits in age and metallicity used in determining the SFH of the Sculptor dSph we consider all information currently available in the literature. We first consider the spectroscopic MDF from \ion{Ca}{ii} triplet and HR spectroscopy. The range in [Fe/H] is limited at the metal-poor end by the availability of the isochrones, which go down to only [Fe/H]$\approx$$-$2.5 dex. This is not a problem, because the spectroscopic MDF shows that the majority of the stars in Sculptor~(92\% of the total) have [Fe/H]$\ge$$-$2.5 dex~\citep{Starkenburg10}. The metal-rich is limited by the absence of stars with [Fe/H]$\ge$$-$1.0 dex in the observed MDF~(see Figure~\ref{fealprel}). A binsize of 0.2 dex is assumed for [Fe/H], similar to the average uncertainty on [Fe/H] in the spectroscopic sample. \\
To constrain [$\alpha$/Fe] we use the 89 stars in the HR sample in the central 25$^{\prime}$ diameter region of Sculptor. Figure~\ref{fealprel} shows that a well-defined [$\alpha$/Fe]$-$[Fe/H] relation from HR spectroscopy is present in the centre of Sculptor. We assume that there is no change in this relation with radius and directly use it to determine the range in [$\alpha$/Fe] that corresponds to the [Fe/H] used in a particular stellar population~(the region defined by the solid black lines in Figure~\ref{fealprel}). In this way we generate a set of populations that cover the full range of [Fe/H] and [$\alpha$/Fe] in the spectroscopic observations. \\
To restrict our choice of possible ages we notice that the distribution of stars in the observed CMDs do not allow any stars $\le$5 Gyr old to be present in Sculptor~\citep{deBoer2011A}. Assuming a maximum age of 14 Gyr, for the age of the Universe, the range of ages we consider is thus between 5 and 14 Gyr old, with a bin size of 1 Gyr.
\begin{figure}[!ht]
\centering
\includegraphics[angle=0, width=0.49\textwidth]{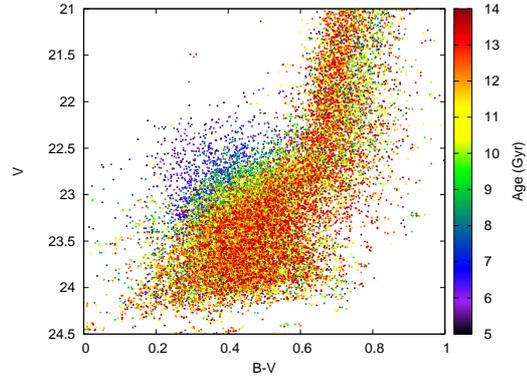}
\caption{Synthetic V,B-V CMDs of the MSTO region of the Sculptor dSph, colour coded by age as inferred from the SFH.  \label{CMDoverlay}} 
\end{figure}

\section{Results}
\label{results}
We have described our method and carried out tests to show that it works as expected using real and synthetic test data~(see Section~\ref{Scltests}) Now we apply Talos to our photometric and spectroscopic data sets of the Sculptor Sph~\citep[][Hill et al., in prep]{deBoer2011A, Starkenburg10}. \\
To derive the SFH of Sculptor the available photometry~(excluding the RGB) is fit simultaneously with the spectroscopically determined MDF. Within~r$_{ell}$$\le$0.43 degrees photometry in three filters~(B,V and I) is used to determine the SFH, while further out only the V and I filters can be used. For radii r$_{ell}$$\le$0.2 degrees the spectroscopic MDF obtained from HR spectroscopy is used. This MDF is used in the SFH fitting for [Fe/H]$\ge$$-$1.9 dex, to avoid the luminosity function bias described in Section~\ref{Sclspecifics}. Below this value only the photometry is used to constrain the MDF and SFH. For r$_{ell}$$\ge$0.2 degrees the MDF obtained from \ion{Ca}{ii} triplet data is used, and no metallicity cut is applied. Furthermore, we include a static background Hess diagram to account for the presence of a BSS population~(as discussed in Section~\ref{Sclspecifics}). The observed HB and AGB are ignored, since the isochrone set we use does not include these evolutionary features. \\
The final Star Formation History and Chemical Evolution History~(CEH) of the entire Sculptor dSph~(out to~r$_{ell}\approx$1~degree) are presented in Figure~\ref{overallSFH}. The SFH and CEH display the rate of star formation at different age and metallicity respectively, in units of solar mass per year or dex respectively, over the range of each bin. The total mass in stars formed in a bin can be obtained by multiplying the star formation rate by the age or metallicity range of that bin. \\
The overall SFH shows that the Sculptor dSph is dominated by an ancient~($>$10 Gyr old), metal poor stellar population. A tail of more metal-rich, younger stars is seen at a lower SFR down to an age of $\approx$6-7 Gyr. The shape of the SFH shows that the star formation rate declines with age, suggesting a single episode of star formation over an extended period of time~($\approx$7 Gyr). 
\begin{figure*}[!htb]
\centering
\includegraphics[angle=0, width=0.33\textwidth]{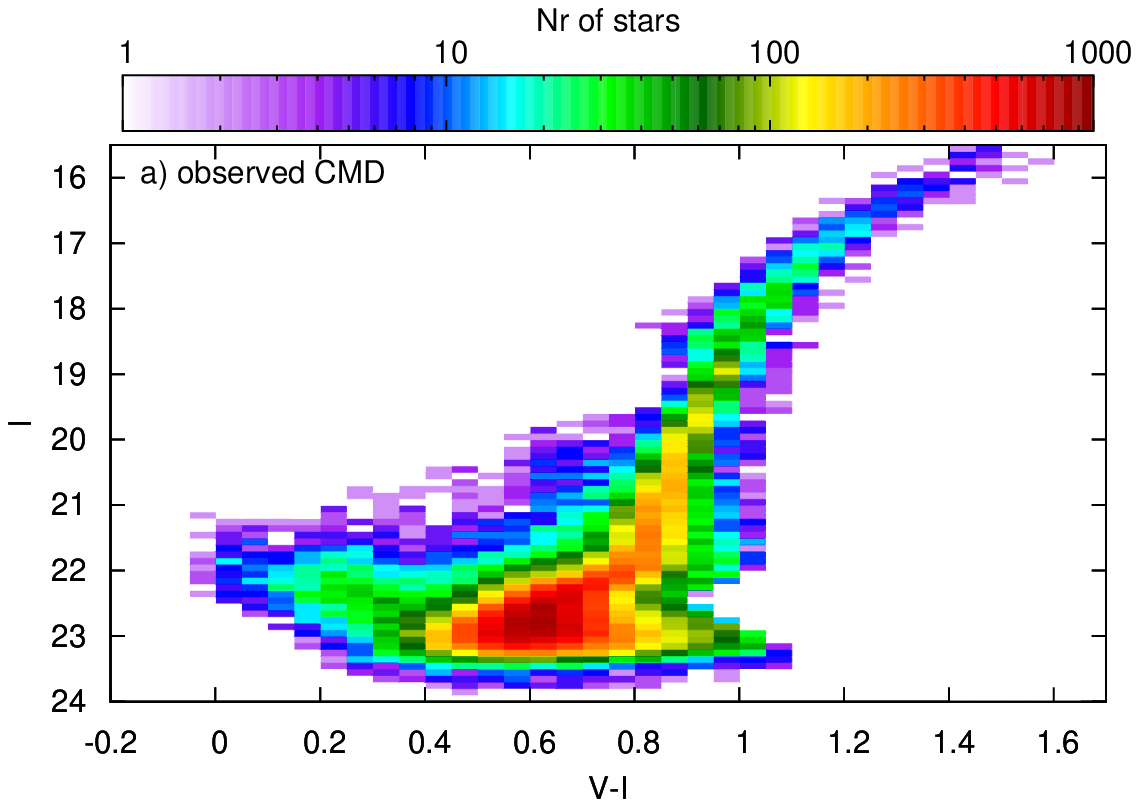}
\includegraphics[angle=0, width=0.33\textwidth]{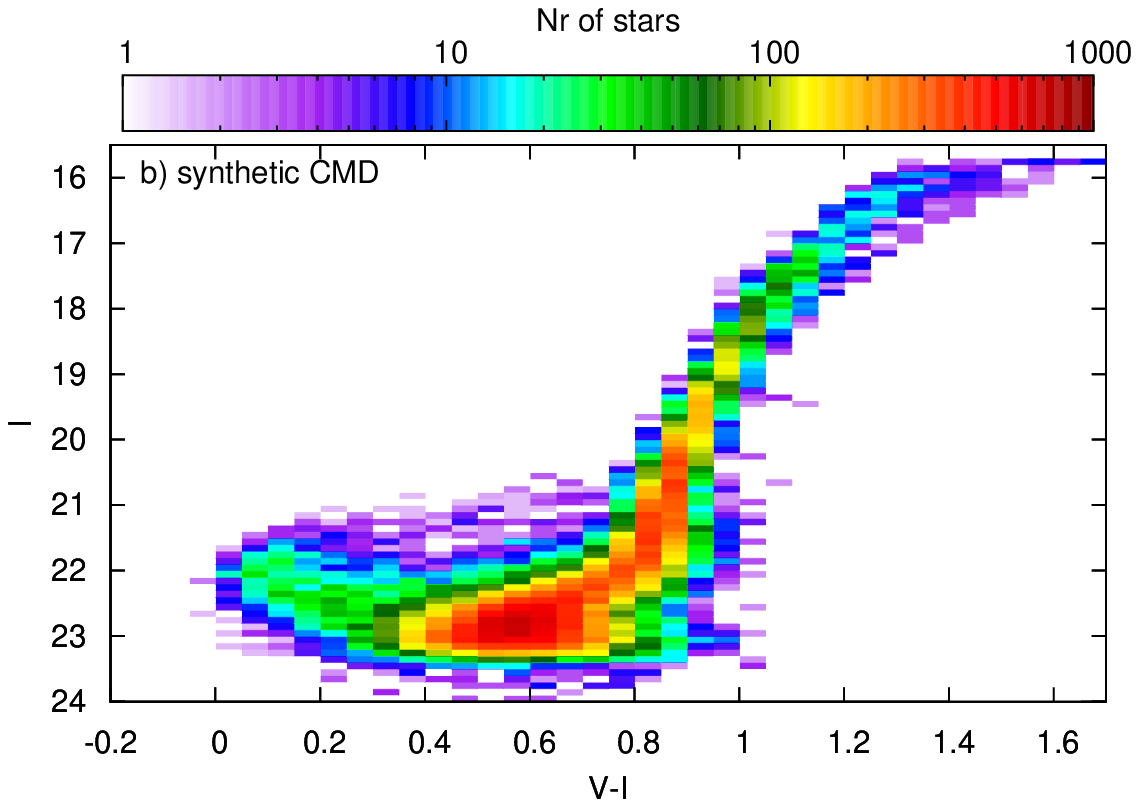}
\includegraphics[angle=0, width=0.33\textwidth]{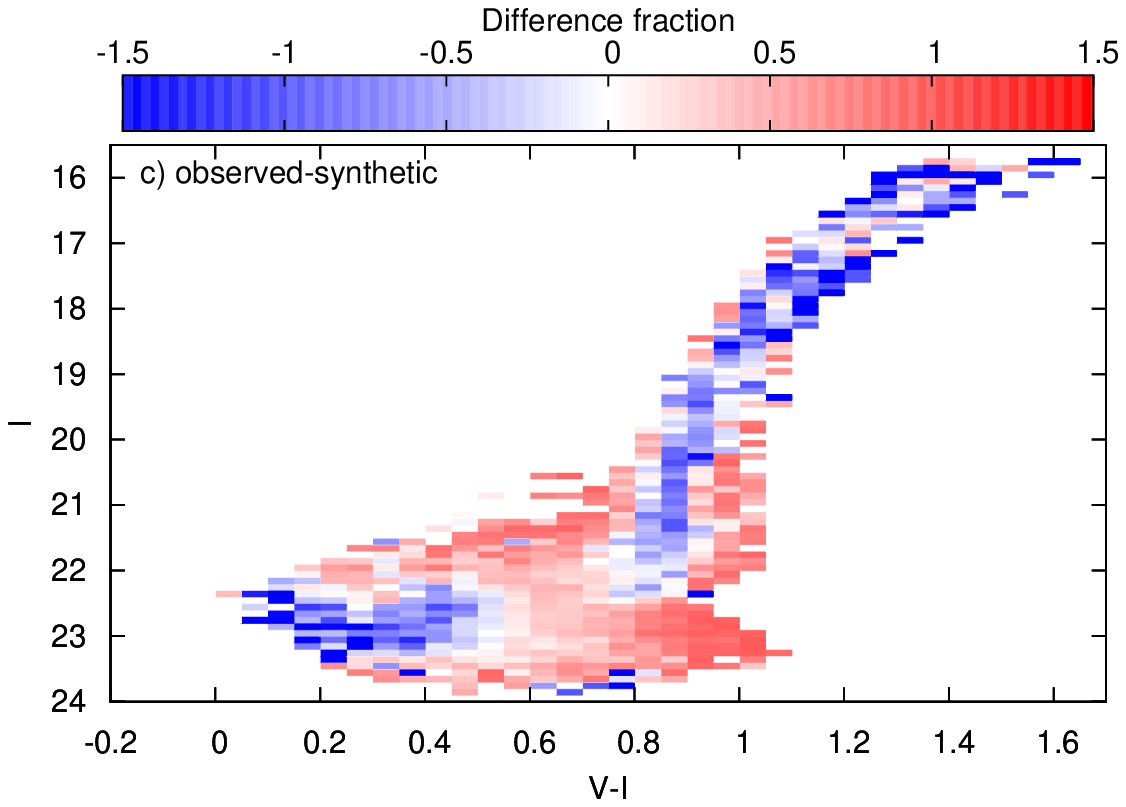}
\caption{\textbf{a)} The observed, cleaned I,V-I CMD of the Sculptor dSph. Colours represent the number of stars per bin, on a logarithmic scale. To allow comparison with the synthetic CMD, the HB feature is removed from the CMD. \textbf{b)} The best synthetic I,V-I CMD of Sculptor, obtained from the SFH determination. The HB and AGB evolutionary features are not included in the isochrone set used, and so they are not present in the synthetic CMD. \textbf{c)} The difference between the observed and synthetic I,V-I CMD of Sculptor, expressed as a fraction of the observed number of stars. \label{SclobsmodHess}} 
\end{figure*}
\\
The SFH is determined using the Darthmouth Stellar Evolution Database~\citep{DartmouthI}. To determine the effect of using different isochrones sets, the SFH of the central part of Sculptor~(r$_{ell}$$\le$0.116 deg) was obtained using three different isochrone sets~\citep{DartmouthI, TeramoI, TeramoII, YonseiYaleI, YonseiYaleII}. The obtained SFHs displayed in Figure~\ref{isotest} show that the obtained solution is very similar in all cases, due to the fact that the SFH determination is dominated by the MSTO region, for which the different isochrone agree at old ages. This gives confidence to adopting the Dartmouth isochrone library for the determination of the final SFH of Sculptor. \\
Using the SFH it is possible to simulate CMDs of the Sculptor dSph, which can be used to show the distribution of age across the CMD in detail. Figure~\ref{CMDoverlay} shows the synthetic V,B-V CMD of the Sculptor MSTOs colour coded with age as inferred from the SFH. The separation of different ages is clearly visible on the MSTO, which highlights the necessity of photometry going down to the oldest MSTOs to obtain accurate ages. 
\begin{figure}[!ht]
\centering
\includegraphics[angle=0, width=0.49\textwidth]{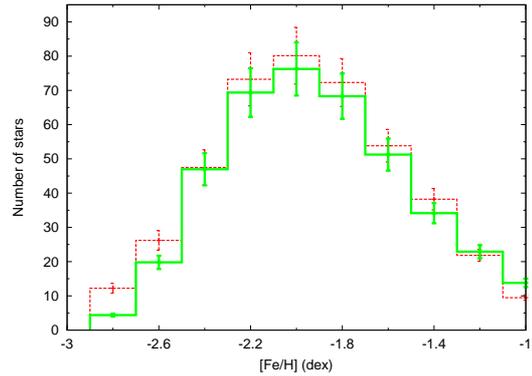}
\caption{The best matching simulated MDF~(green) solid histogram and observed MDF~(red) dashed histogram from spectroscopic observations. \label{MDFfull}} 
\end{figure}

\subsection{Reliability of the SFH}
The most basic check of the reliability of the recovered SFH is to compare the synthetic and observed CMDs. Figure~\ref{SclobsmodHess} shows the observed I,V-I CMD of the Sculptor dSph and the synthetic CMD corresponding to the best matching SFH. Furthermore, the difference between both CMDs is shown, expressed as a fraction of the observed star counts in each bin.\\
The total number of stars in the synthetic CMD is consistent with the observed CMD to within a few percent, showing that the total mass in stars is well matched to the observations. The CMDs are in general a good match, over most of the MSTO region. The static background BSS is not reproduced with the same colour distribution as the observed stars, leading to an under-dense region at blue colours~(V$-$I$\approx$0.2) in Figure~\ref{SclobsmodHess}c. However, the transition from the BSS into the MSTO region is matched well, despite being modelled in a simple way~(See Section~\ref{Sclspecifics}). Figure~\ref{SclobsmodHess}c shows that the colours of the RGB extend to redder colours than observed in Sculptor. The RGB colours of the adopted isochrone set are slightly too red for the metallicity and age determined from the MSTO region and~\ion{Ca}{ii} triplet spectroscopy, which is a known problem for theoretical isochrones~\citep[e.g.,][]{Gallart05}. The HB and AGB are obviously not reproduced in the synthetic CMD, as they are not included in the Dartmouth isochrone set. \\
On the RGB a comparison can be made between the observed spectroscopic MDF and the MDF obtained from the SFH determination, as shown in Figure~\ref{MDFfull} for the full extent of the Sculptor dSph. We can see that the synthetic MDF is mostly consistent with the observed MDF within the errors, as expected, given that the MDF is used as an input in the SFH determination. A population which is not so well fit is the tail of metal-poor stars~([Fe/H]$<$-2.5 dex). From the spectroscopic observations it is known that these stars are present in Sculptor, although in small numbers~\citep[$\approx$8\%, see][]{Starkenburg10}. The models used to fit the SFH do not contain metallicities low enough to include these stars. Despite these uncertainties we feel confident that the differences between the observed and synthetic CMDs and MDFs are sufficiently low to conclude that the fitted SFH constitutes an accurate representation of the full range of ages of stellar populations present in Sculptor. \\
\begin{figure*}[!ht]
\centering
\includegraphics[angle=0, width=0.97\textwidth]{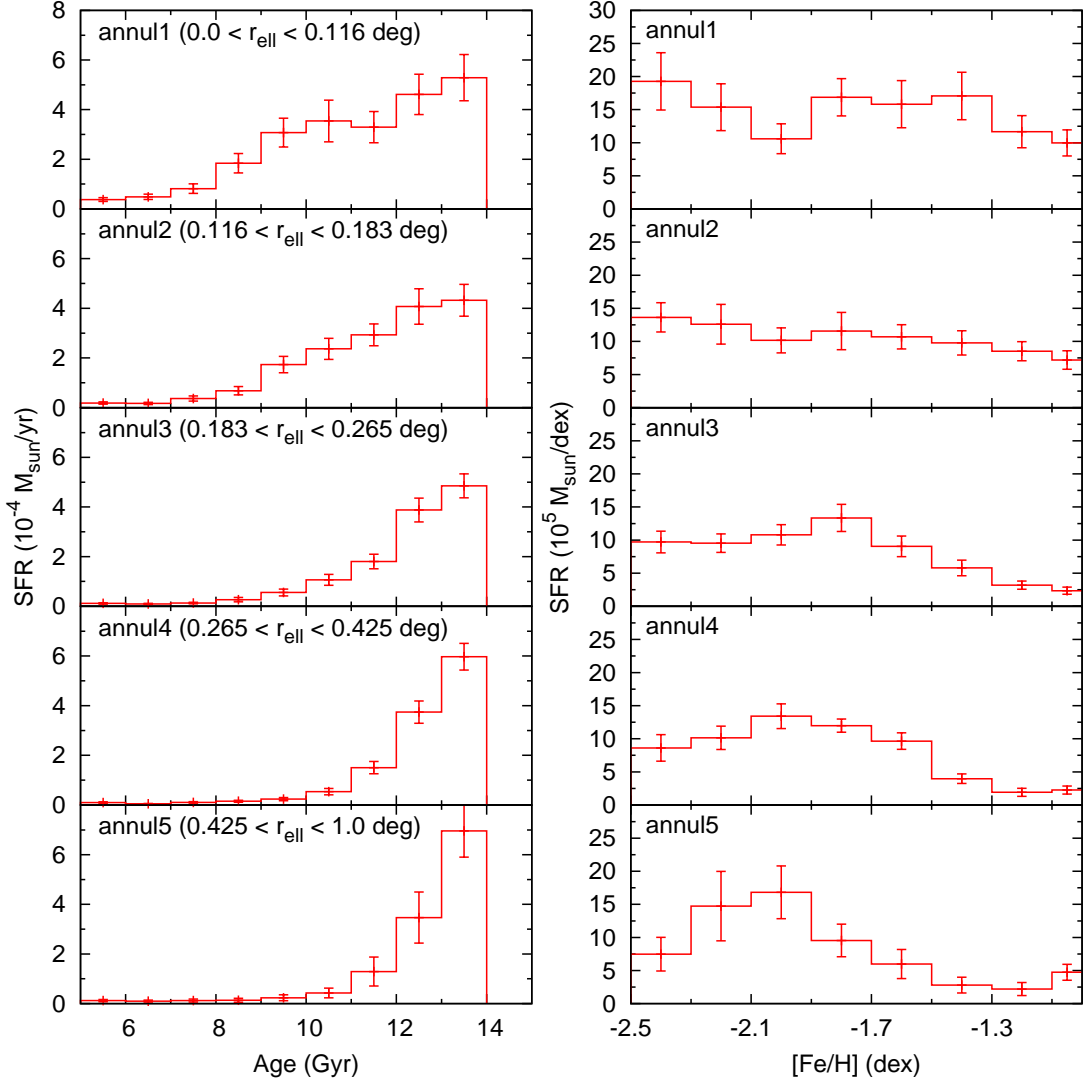}
\caption{The SFH and CEH~(obtained by fitting all available photometry and spectroscopy) for each of the 5 annuli of Sculptor, with the corresponding radial extent indicated in the plot. For annulus 5 only the I,V-I CMD is used together with the spectroscopy to constrain the SFH. \label{spatialSFH}} 
\end{figure*}
\begin{figure}[!ht]
\centering
\includegraphics[angle=0, width=0.49\textwidth]{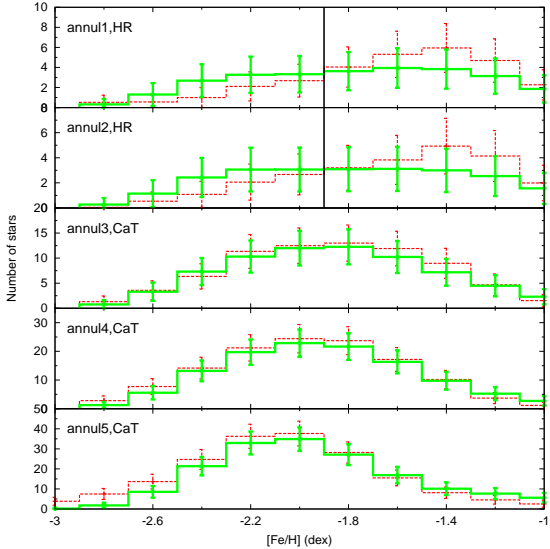}
\caption{The observed~(red) dashed histogram and the synthetic~(green) solid histogram MDFs from the SFH analysis in the five annuli. The solid~(black) line in the upper two panels shows the [Fe/H] above which the HR spectroscopic MDF is unbiased, and can be used in the SFH determination. \label{MDFall}} 
\end{figure}

\subsection{Spatial variations in the SFH}
\label{differentregions}
Due to the extensive spatial coverage of both the MSTO photometry and RGB spectroscopy we are able to determine the SFH over most of the area of the Sculptor dSph. Thus, it is possible to determine the radial variation in the SFH in 5 annuli~(see Figure~\ref{spatialSFH}), each containing a similar number of stars observed in the V and I filters. For comparison, the Sculptor core radius is 0.1 degrees, just within the first annulus. The V and I filters were chosen, since they are offer the most complete spatial coverage further from the centre. 
\begin{figure}[!t]
\centering
\includegraphics[angle=0, width=0.49\textwidth]{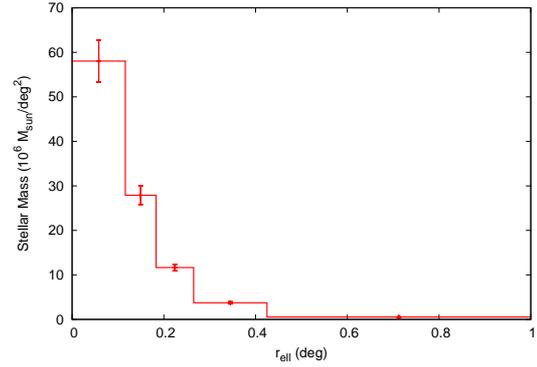}
\caption{The radial distribution of the total stellar mass per square degree in the Sculptor dSph, as determined from the SFH modelling. This provides the total mass of stars formed over the lifetime of Sculptor, 7.8$\times$10$^{6}$ M$_{\odot}$. \label{radtotal}} 
\end{figure}
\\
The SFH is determined independently in each of these annuli in the same way as for the full galaxy. The SFH and CEH of the five selected spatial regions are shown in Figure~\ref{spatialSFH}, using all photometric filters as well as the available spectroscopy. In the outermost annulus the SFH is determined only using the I,V-I CMD, together with the spectroscopic MDF. Figure~\ref{spatialSFH} shows that the SFH changes significantly with distance from the centre of Sculptor. The metallicity and age gradient discussed by~\citet{deBoer2011A} is quantified here. The younger, metal-rich populations are found mostly in the central region and drop off towards the outer parts.  \\
Figure~\ref{MDFall} shows the observed and synthetic MDFs resulting from the SFH determination for all five annuli in Sculptor. The MDF shown in the upper two panels comes from HR spectroscopy, while the MDF in the lower three panels comes from \ion{Ca}{ii} triplet spectroscopy. Figure~\ref{MDFall} shows that the synthetic MDFs resulting from the SFH determination are consistent with the observed MDFs within the errors. In the upper two panels the synthetic MDF lies above the observed MDF for metallicities lower than indicated by the black line and below the observed MDF for higher metallicities. This is similar to the effects of the luminosity function bias seen in Figure~\ref{lumfuncfrac}.  
\begin{figure}[!ht]
\centering
\includegraphics[angle=0, width=0.49\textwidth]{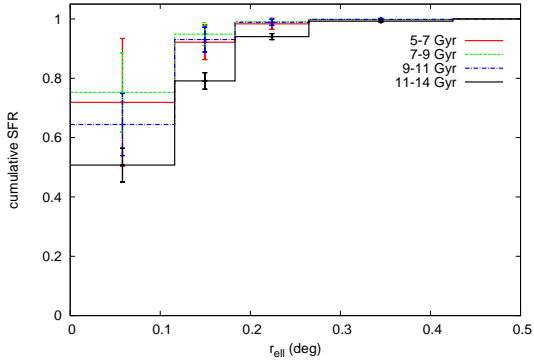}
\caption{The cumulative radial distributions of stellar populations with different ages. \label{radvariations}} 
\end{figure}
\\
Using the SFH at different positions it is possible to determine the radial distribution of the total mass of stars in Sculptor. This is done by integrating the SFRs over the duration of star formation, at different radii from the centre. Figure~\ref{radtotal} shows the radial distribution of the total stellar mass, obtained from the SFH modelling. The total mass in stars formed is highest in the central parts of Sculptor~(r$_{ell}$$\le$0.2 degrees) and drops off steeply towards the outskirts. The total mass in stars formed over all radii in Sculptor for our SFH is 7.8$\times$10$^{6}$ M$_{\odot}$ within an elliptical radius of 1 degree or 1.5 kpc. To determine the core radius of the total mass distribution we use a Sersic profile to fit the distribution shown in Figure~\ref{radtotal}. The resulting core radius r$_{c}$=0.11$\pm$0.04 degrees, is consistent within the errors with that derived from the observed density profile~\citep{Battaglia081}.\\
It is also possible to quantify the radial distribution of populations with different ages. Figure~\ref{radvariations} shows the cumulative radial distributions of stellar populations with different ages, as determined from SFH modelling. Figure~\ref{radvariations} is another way of showing that the oldest population is the most extended component, while younger populations show an increasing central concentration, as qualitatively described in~\citet{deBoer2011A}. 

\subsection{Resolving bursty star formation}
\label{burstystarformation}
\begin{figure*}[!t]
\centering
\includegraphics[angle=0, width=0.49\textwidth]{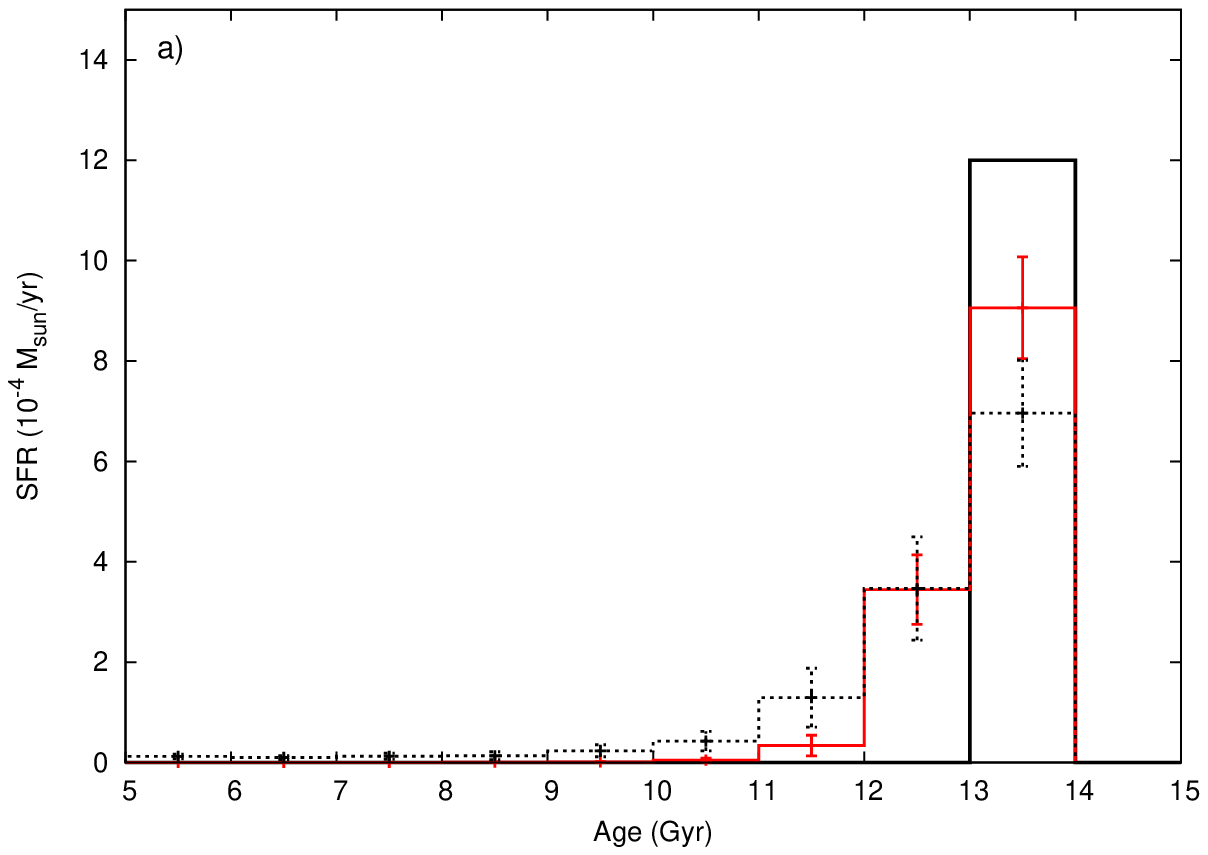}
\includegraphics[angle=0, width=0.49\textwidth]{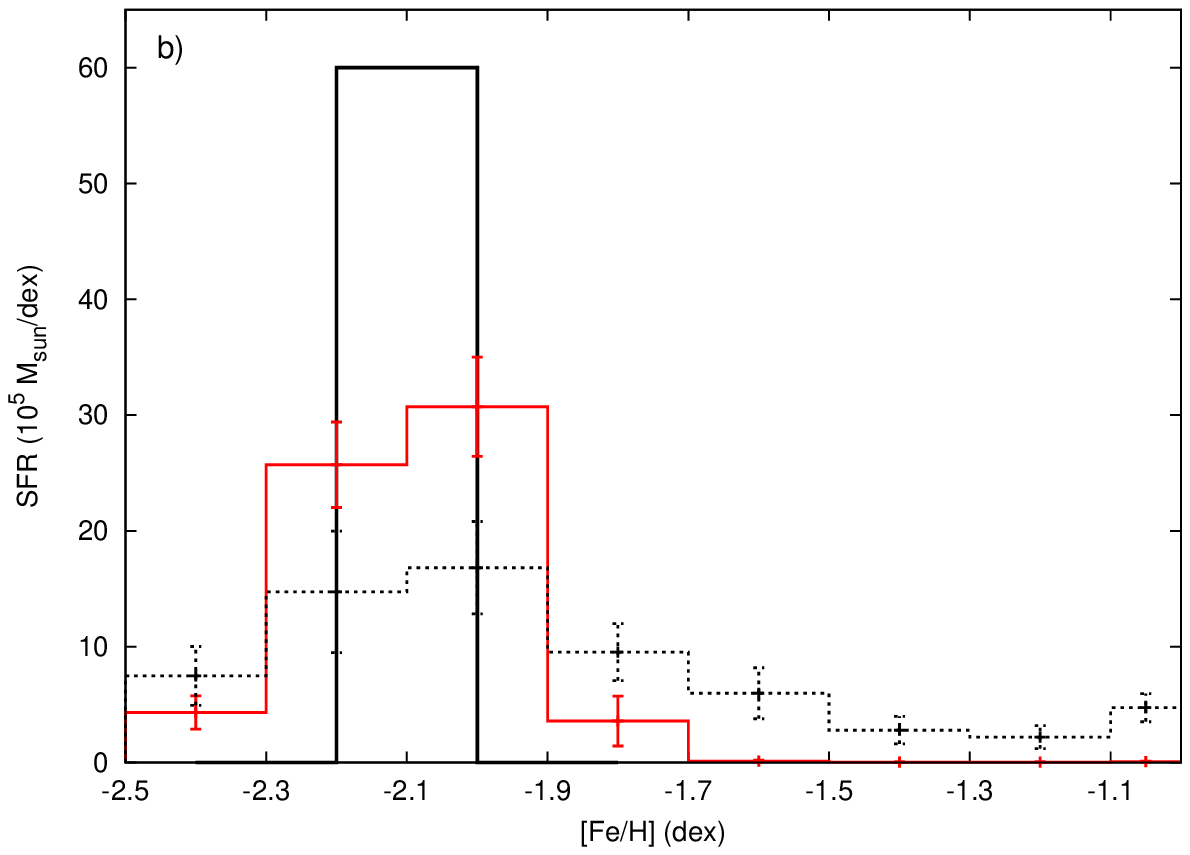}
\caption{The recovered and input~(\textbf{a}) SFH and~(\textbf{b}) CEH of a synthetic population with a single short~(10 Myr) episode of star formation at an age of 13.5 Gyr. The solid black histogram shows the input population given the SFH binning, while the red histogram shows the recovered SFH. For comparison, the adopted SFH of annulus 5 of Sculptor is shown as the black, dashed histogram. \label{burst1}} 
\end{figure*}
\begin{figure}[!htb]
\centering
\includegraphics[angle=0, width=0.49\textwidth]{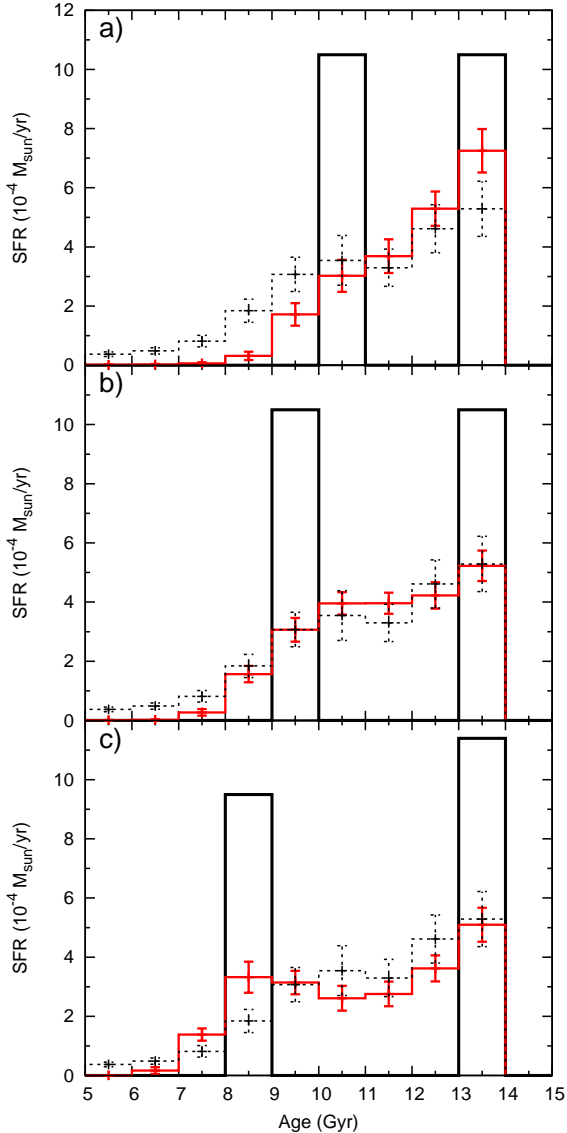}
\caption{The input and recovered SFH for a series of synthetic populations with two short~(10 Myr) bursts of star formation. The oldest takes place at an age of 13.5 Gyr, with the second burst at an age of~(\textbf{a}) 10.5 Gyr,~(\textbf{b}) 9.5 Gyr and~(\textbf{c}) 8.5 Gyr.  The solid black histogram shows the input population, given the SFH binning adopted. The solid red histogram shows the recovered SFH along with statistical errorbars. For comparison, the adopted SFH of annulus 1 of Sculptor~(Figure~\ref{spatialSFH}) is shown as the black dashed histogram. \label{burst2}} 
\end{figure}

One crucial aspect of the interpretation of our SFH is: can we distinguish between bursty and continuous star formation, given our SFH resolution and uncertainties? The intrinsic uncertainties on the SFH determination~(as described in Section~\ref{uncertainties}) result in a smoothing of the star formation rates, potentially hiding the presence of bursts of star formation. \\ 
To test our ability to distinguish a series of distinct bursts of star formation from continuous star formation in Sculptor, we carry out a series of experiments using the final SFH of the innermost and outermost annuli of Sculptor~(anull1 and annul5 in Figure~\ref{spatialSFH}). The metallicities and star formation rates from the final SFH are adopted, and distributed over one or more short bursts of star formation. \\
For a single narrow~(10 Myr) burst of star formation at an age of 13.5 Gyr and $-$2.2$\le$[Fe/H]$\le$$-$2.0 dex metallicity range the recovered and input SFH and CEH for annulus~5 are shown in Figure~\ref{burst1}. The recovered SFH matches well with the final adopted SFH of the outermost annulus given in Section~\ref{differentregions}. This suggests that the star formation in the outermost part of Sculptor can be approximated with a single, short burst of star formation. However, the SFH and CEH shown in Figure~\ref{burst1} indicate that the spread in age and metallicity of the burst must be more extended than assumed here. \\
A similar analysis for annulus~1 of Sculptor shows that the recovered SFH using a single, short burst is clearly inconsistent with the adopted SFH of the inner annulus. A comparison of the synthetic CMD with the observed CMD of annulus~1 shows a bad fit to the data, with a value of $\chi^{2}$=5.51~(compared to $\chi^{2}$=2.94 for the adopted SFH of annul1), making it unlikely to be an accurate representation of the actual SFH in the centre of Sculptor. Additionally, it seems unphysical to assume that the large spread in metallicity observed in the centre of Sculptor~(1.5 dex) can be built up during a single short episode of star formation. \\
We continue by adding a second burst of star formation to see if this can improve the match to the SFH of the innermost annulus of Sculptor. Figure~\ref{burst2} shows the recovered SFH of a synthetic population with two narrow~10Myr bursts of star formation at different age, with metallicities and star formation rates determined from the adopted SFH of annulus~1. In each test an old burst was assumed to take place at an age of 13.5 Gyr, with an additional burst at a younger age. Figures~\ref{burst2}a,b show that the input SFH is not recovered as two individual bursts if the age separation between the two bursts is less than 4 Gyr. Only when the youngest burst is placed at an age of 8.5 Gyr, that a separation can be seen between the two bursts in the SFH, as shown in Figure~\ref{burst2}c. 
\\
The SFHs shown in Figure~\ref{burst2} are in relatively good agreement with the final adopted SFH of annulus~1, especially when the youngest burst is placed at 9.5 Gyr~(Figure~\ref{burst2}b). The recovered SFH shows features consistent with the adopted SFH at similar strengths. However, a comparison of the synthetic CMD to the observed CMD of annulus~1 shows that the synthetic populations provide a bad fit to the observed CMD, with $\chi^{2}$ values of 4.32, 4.10 and 4.28 respectively for the populations with the young burst at 10.5, 9.5 and 8.5 Gyr. The smooth SFH, as given in Section~\ref{differentregions} provides a better representation of the observed CMD, which makes it unlikely that strong, short bursts of star formation dominated the SFH of the Sculptor dSph.\\
A synthetic population with three narrow~10Myr bursts of star formation, at 13.5, 10.5 and 8.5 Gyr, gives a better agreement with the observed SFH. Furthermore, the synthetic CMD provides a better fit to the observed CMD of annulus~1, with a $\chi^{2}$ value of 3.96, although still not as good as the smooth SFH given in Section~\ref{differentregions}. Taking this approach further, adding ever more short bursts of star formation will provide a better representation of the observed CMD, to the point where we can not distinguish between bursty and continuous star formation. In the end, this is how our model builds up the SFH. The limits of the bursty nature of the Sculptor SFH cannot be tied down with CMD analysis at such old ages as these with better accuracy than shown in Figure~\ref{burst2}. However, the form of the CMD and the metallicities of individual RGB stars would be markedly different if star formation had truly occurred in a few short bursts, and also other evidence such as the broad MDF makes it unlikely.  

\subsection{The timescale for chemical evolution}
\label{indivages}
\subsubsection{Determining ages of individual stars}
Using the SFH determined for Sculptor it is possible to go back to the spectroscopic sample and determine the ages of the individual RGB stars. For each observed star with spectroscopic abundances we find all the stars in a synthetic CMD made with the SFH of Sculptor, with the same magnitude~(in all filters) and metallicity within the observed uncertainties. These stars are considered to be representative of the age of the observed star. The mean age of the matched sample is adopted as the age of the observed star, with the corresponding standard deviation of the sample as an errorbar. In the case of Sculptor around 100 synthetic stars are typically available to compute the mean and standard deviation for each observed star, ensuring enough stars are present to determine a reasonable errorbar.\\
To demonstrate the advantage of using the SFH results to obtain ages for individual RGB stars a comparison is made to the standard simple isochrone fitting technique, which uses a fine grid of [Fe/H], age and [$\alpha$/Fe] to obtain the age and uncertainty for an observed star along an isochrone. Figure~\ref{AMR}a shows the Age-Metallicity Relation~(AMR) obtained using simple isochrone fitting, and Figure~\ref{AMR}b using our full SFH approach. The AMR determined using isochrone fitting has bigger age uncertainties and allows younger ages for each star, which are inconsistent with the SFH~(see Fig~\ref{overallSFH}). This is because the age for simple isochrone fitting is dependent only on the colour of the RGB and the measured [Fe/H], which allows a wide range in ages for each observed star. Conversely, the more constrained AMR coming from the SFH is because information from the entire CMD~(and all modelled evolutionary phases) is used to constrain the age of a single observed star. 

\subsubsection{Chemical evolution of the Sculptor dSph}
With accurate ages assigned to each RGB star in the HR spectroscopic sample it is possible to effectively measure the evolution of particular elements and, for the first time, directly determine timescales for chemical evolution. Figure~\ref{MgFeage}a shows [Mg/Fe] for RGB stars in Sculptor measured from HR spectroscopy, with the age of the individual stars colour coded. Figure~\ref{MgFeage}b shows that [Mg/Fe] follows a uniform trend with age, such that the oldest stars are more enhanced than the younger. We can also determine the age at which the ``knee"~\citep{Tinsley79, Gilmore91} occurs. This marks the time at which SNe Ia start to contribute to the [Fe/H] content of a galaxy~\citep{Matteucci90,Matteucci03}. In Sculptor, the turnover takes place between $-$1.8$<$[Fe/H]$<$$-$2.0 dex, from which we determine that the [Fe/H] from SNe Ia started to be produced 10.9$\pm$1 Gyr ago. Therefore, the SNe Ia started contributing noticeably to the chemical evolution of Sculptor approximately 2$\pm$1 Gyr after the beginning of star formation. This is the first direct measure of this timescale, although it has previously been inferred from SNe Ia timescales and chemical evolution models~\citep[][and references therein]{Mannucci06}.

\section{Conclusions}
\label{conclusions}
\begin{figure*}[!htb]
\centering
\includegraphics[angle=0, width=0.49\textwidth]{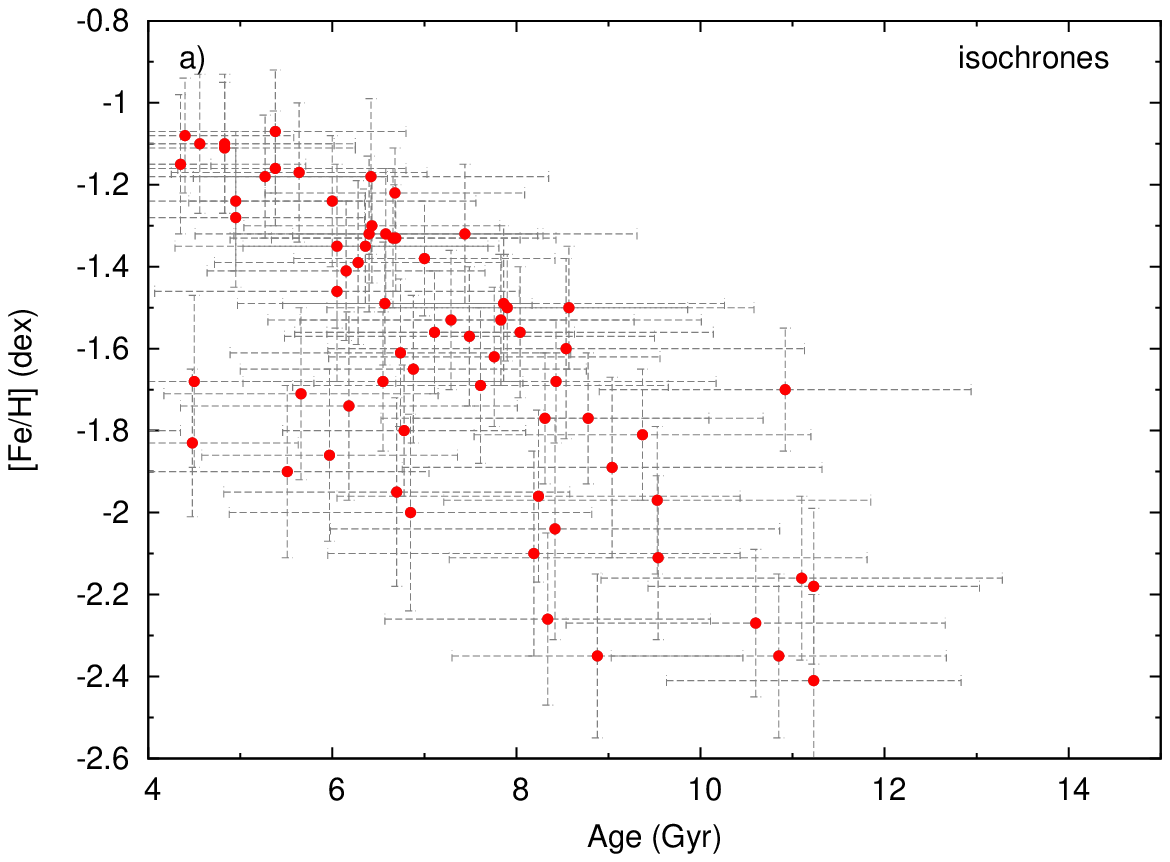}
\includegraphics[angle=0, width=0.49\textwidth]{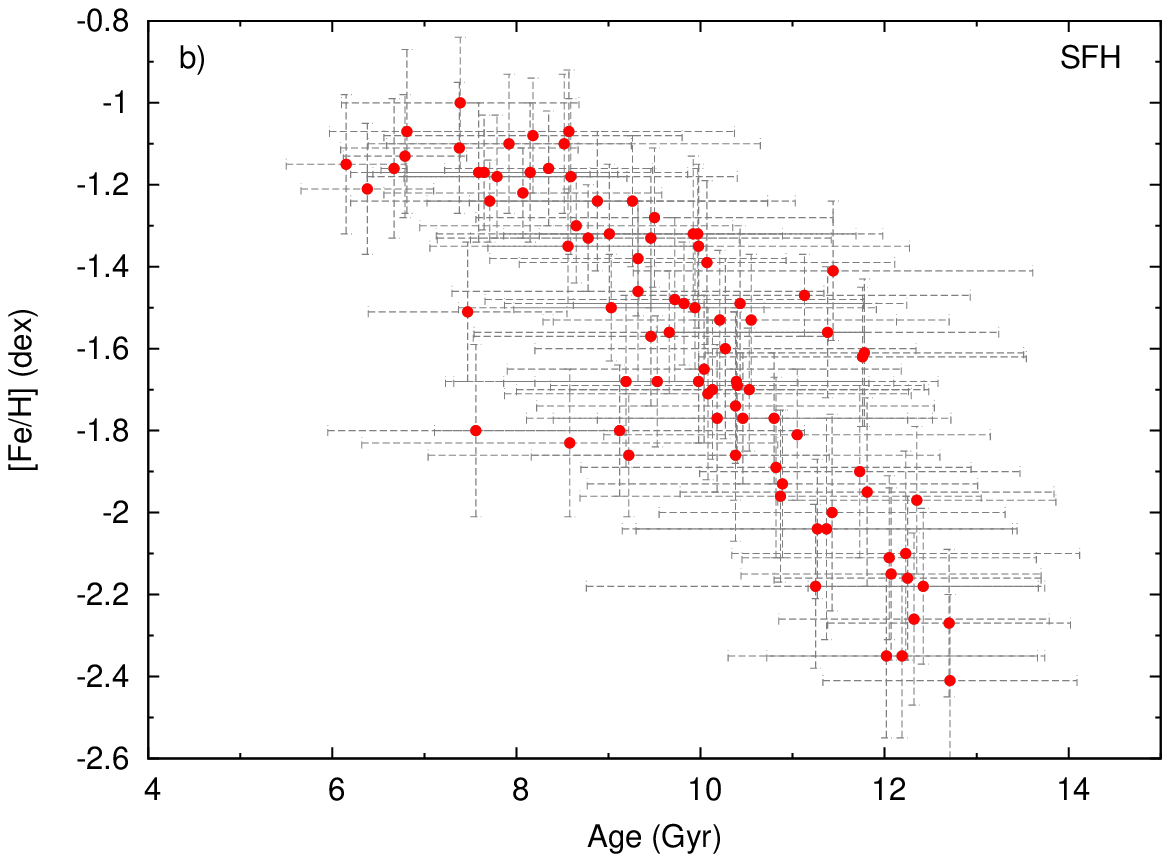}
\caption{The Age-Metallicity Relation of stars on the upper RGB in Sculptor, as determined using~(\textbf{a}) simple isochrone fitting and~(\textbf{b}) from our method incorporating the full SFH and MDF information. \label{AMR}} 
\end{figure*}

\begin{figure*}[!htb]
\centering
\includegraphics[angle=0, width=0.49\textwidth]{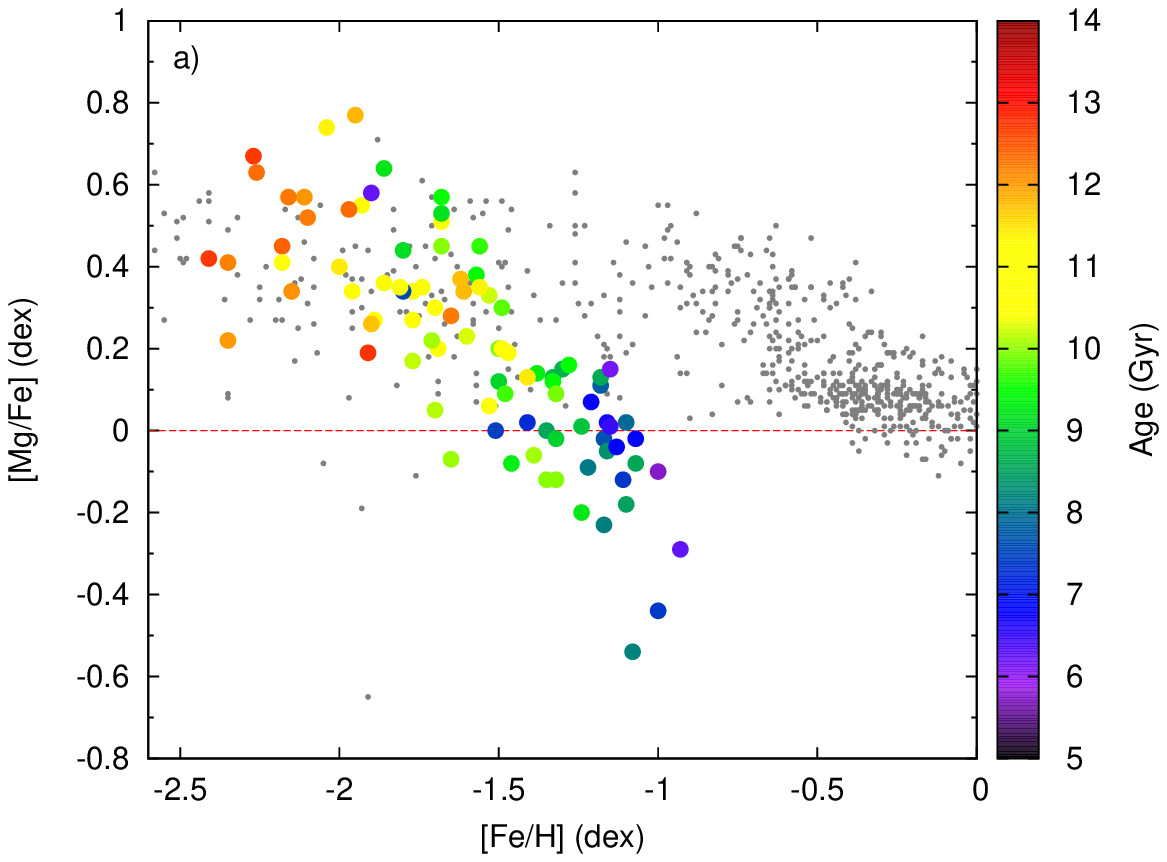}
\includegraphics[angle=0, width=0.49\textwidth]{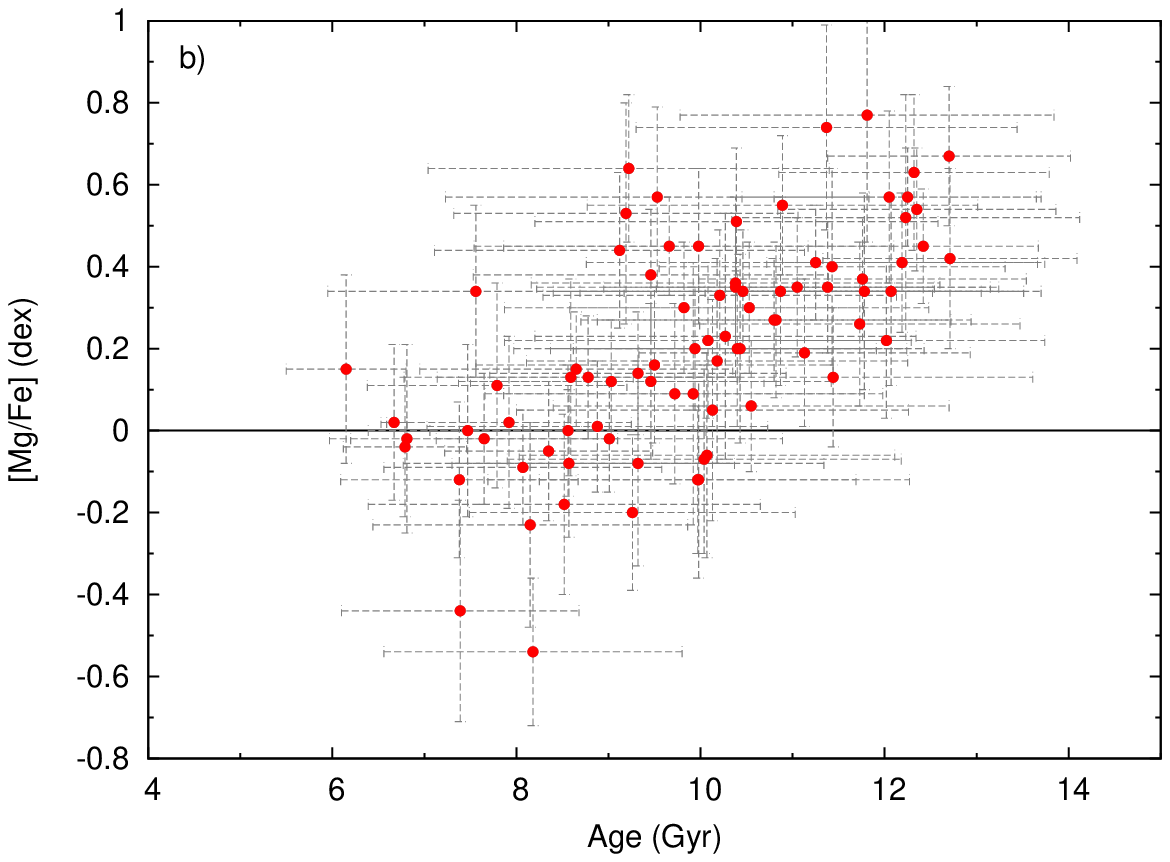}
\caption{\textbf{a)} [Mg/Fe] measurements for the HR spectroscopic sample of RGB stars in the Sculptor dSph~(coloured filled circles). The colours represent the age in Gyr, derived from the SFH. Stars in the Milky Way are shown for comparison~(small grey points). \textbf{b)} [Mg/Fe] is plotted directly against age for the same sample. \label{MgFeage}} 
\end{figure*}

We have presented the first detailed SFH of the Sculptor dSph, using a combination of deep, wide-field multi-colour photometry and spectroscopic metallicities and abundances. The method used to determine the SFH~(described in Section~\ref{method}) directly combines, for the first time, photometry with spectroscopic abundances. \\
The SFH~(see Figure~\ref{overallSFH}) shows features similar to previous rough SFHs~\citep[e.g.][]{DaCosta84,HurleyKeller99,Dolphin02,Tolstoy03}, but resolves the stellar ages with much greater accuracy. Additionally, the SFH quantifies the age and metallicity uncertainties and provides well motivated errors. The SFH and CEH have been determined over a large fraction of the Sculptor dSph~($\approx$80\% of the tidal radius), allowing us to quantify the radial age and metallicity gradients in Sculptor. \\
The SFH shows that star formation took place in Sculptor for an extended period of time, from 14 to 7 Gyr ago~(see Figure~\ref{overallSFH}) with a simple, steadily decreasing trend. The spatially resolved SFH over the whole galaxy~(see Figure~\ref{spatialSFH}) shows that a radial gradient is present in age and metallicity. The MDF~(right-hand panels of Figure~\ref{spatialSFH}) shows that the metal-poor, old populations are present at all radii while the more metal-rich, younger populations are found more toward the centre, consistent with previous qualitative results~\citep{deBoer2011A}. \\
We explored the temporal resolution of our final SFH for the innermost and outermost parts of Sculptor under study. We find that the outermost annulus is roughly consistent with a single episode of star formation. In the inner parts of Sculptor it is difficult to find a bursty SFH that matches all the available data. Therefore, we find no reason to assume bursty star formation episodes here and prefer the overall picture of a single, extended episode of star formation. Additionally, the spatially resolved SFH and CEH are more consistent with an age gradient than with two separate populations. \\
The SFH and CEH at different radii from the centre are consistent with the scenario where Sculptor first experienced a single sizeable burst of star formation at early times, with more metal-rich populations forming ever more concentrated towards the centre until about 6$-$7~Gyr ago, when the star formation activity ceased. \\
Likewise, the chemical evolution of the Sculptor dSph seems to be straightforward, according to Figure~\ref{MgFeage}, which shows that [Mg/Fe] decreases steadily with time. The simple decline in the [Mg/Fe] distribution suggests that the chemical enrichment occurs uniformly over the SFH of the Sculptor dSph, with a change in slope when the SNe Ia start to contribute. \\
The timescale on which SNe Ia start to contribute significantly to the chemical evolution of Sculptor~(approximately 2$\pm$1 Gyr after the beginning of star formation) is comparable within the errors to the timescale expected from the theory of SNe Ia timescales, although inconsistent with timescales of prompt SNe Ia explosions~\citep[see~e.g.][]{Raiteri96,Matteucci01}. The small number of stars and the large scatter on the knee position in Figure~\ref{MgFeage}a means that the turnover age is not exactly defined. A larger sample of metal-poor stars is needed to obtain better statistics. \\
The formation of a well defined, narrow $\alpha$-element vs. metallicity distribution and a tight Age-Metallicity relation~(see Figure~\ref{AMR}b) is consistent with the overall picture that the Sculptor dSph formed stars uninterrupted for an extended period of time. The temporal evolution of individual stars suggests a slow build up over several gigayears. A galaxy with a shorter period of star formation is unlikely to show such a well-defined extended AMR. If the SFH were more bursty, a greater spread in [$\alpha$/Fe] would also be expected, instead of a smooth trend with age and [Fe/H]. \\
As a result, Sculptor can be considered as a good benchmark for isolated star formation over an extended period of time at the earliest epochs. 

\section{Acknowledgements}
\label{acknowledgements}
The authors thank ISSI (Bern) for support of the team ``Defining the full life-cycle of dwarf galaxy evolution: the Local Universe as a template". T.d.B., E.T., E.S. and B.L. gratefully acknowledge the Netherlands Foundation for Scientific Research (NWO) for financial support through a VICI grant. The research leading to these results has received funding from the European Union Seventh Framework Programme (FP7/2007-2013) under grant agreement number PIEF-GA-2010-274151. The authors would like to thank the anonymous referee for his/her comments, that helped to improve the paper.

\bibliographystyle{aa}
\bibliography{references}

\clearpage
\begin{appendix}
\onecolumn
\section{Tests of the method}
\label{tests}
In order to test the ability of Talos to accurately recover a SFH, a number of tests were made on CMDs for which the properties are known, as well as on observed data for a well studied globular cluster.

\subsection{Synthetic tests}
\label{synthetictests}
First of all, a number of simple artificial stellar populations were generated. By recovering the SFH of these populations it is possible to test how accurate our method can recover an input SFH. 
\begin{figure*}[!hbt]
\centering
\includegraphics[angle=0, width=0.49\textwidth]{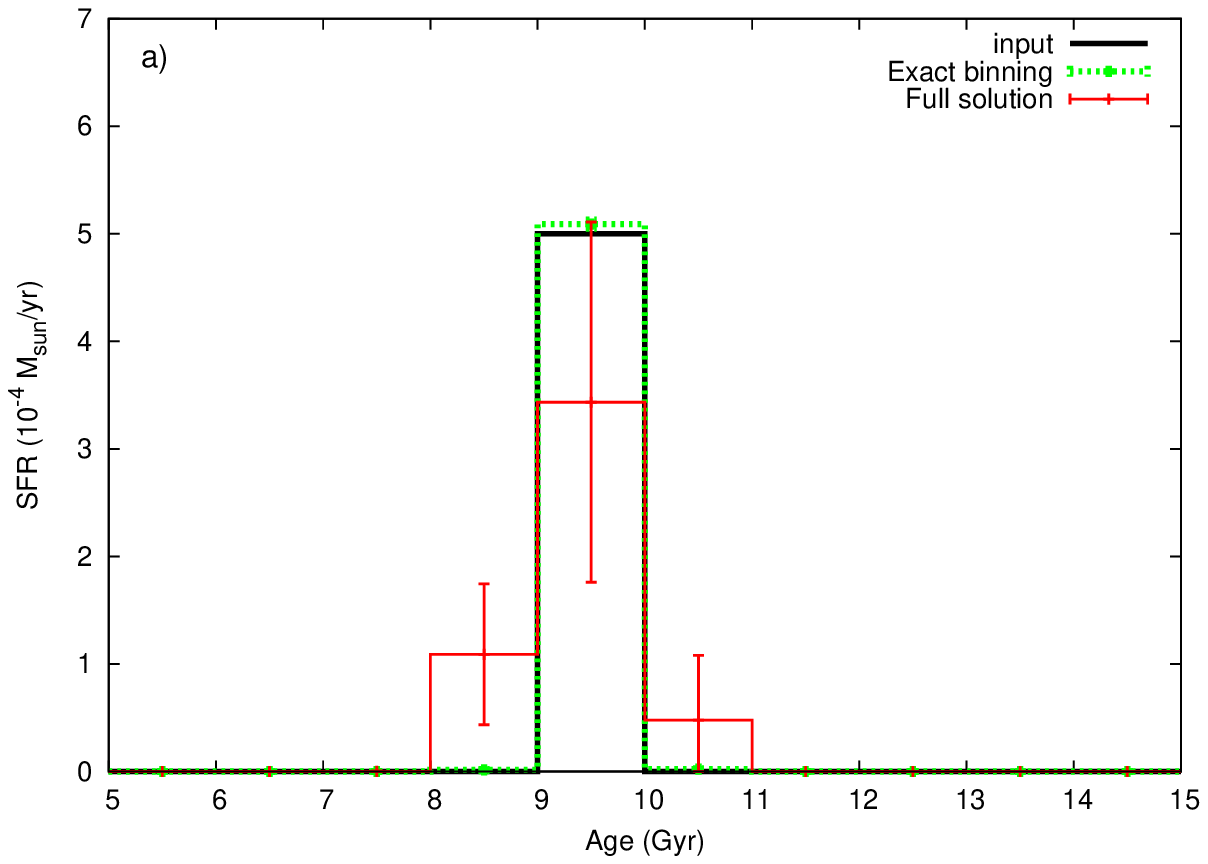}
\includegraphics[angle=0, width=0.49\textwidth]{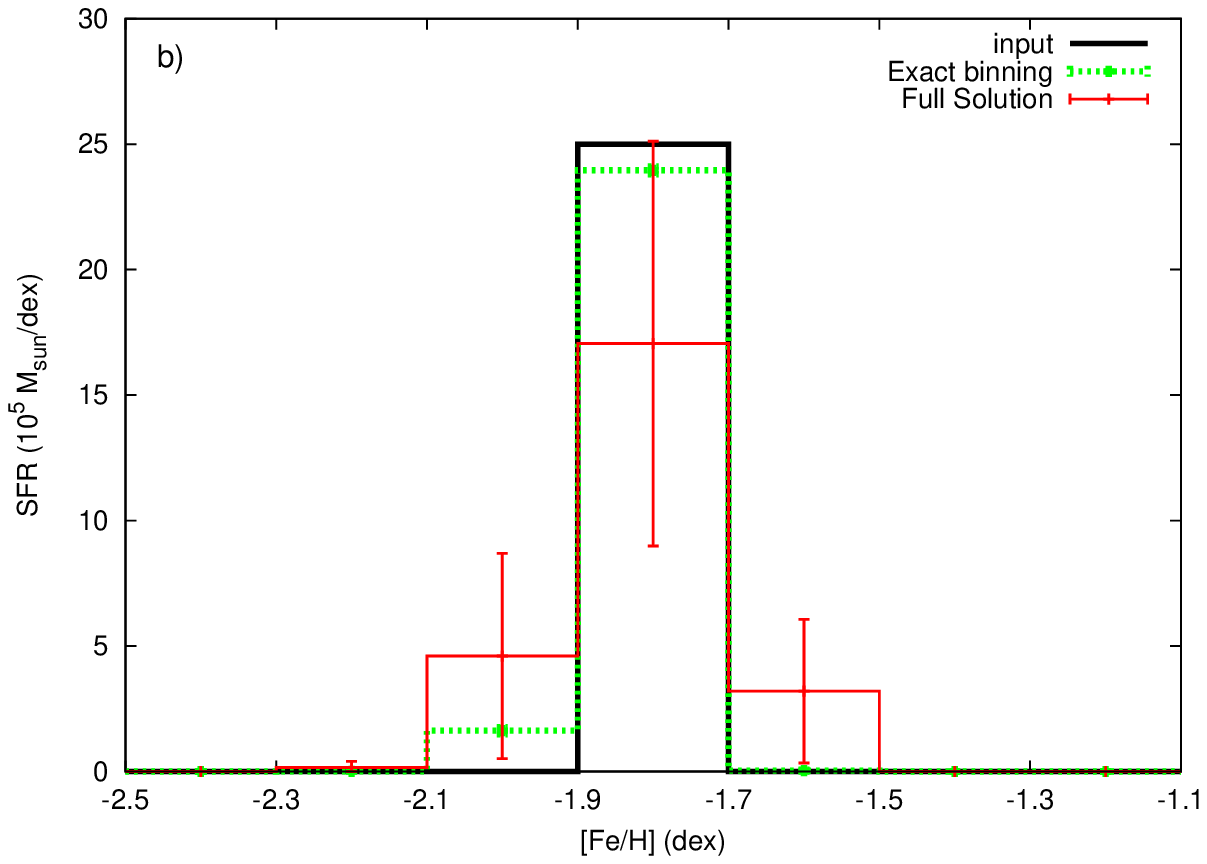}
\includegraphics[angle=0, width=0.49\textwidth]{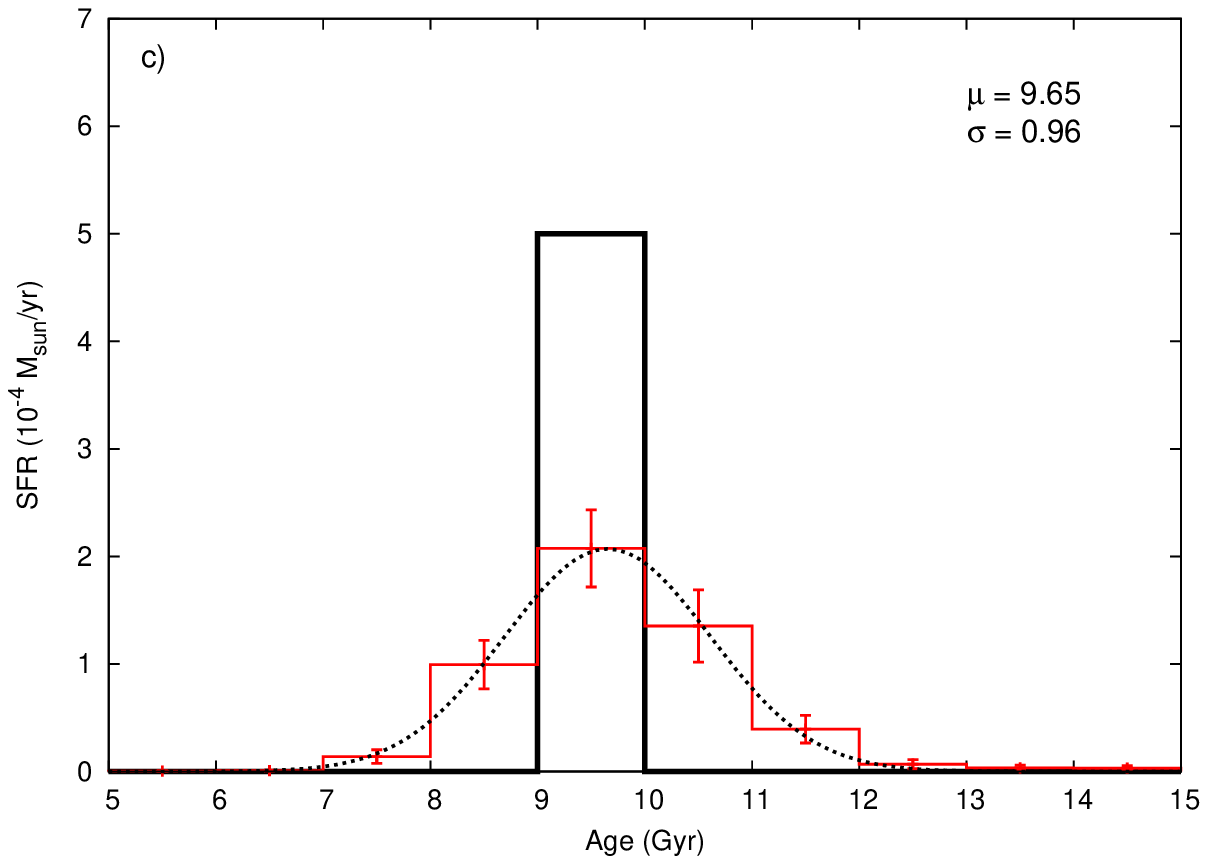}
\includegraphics[angle=0, width=0.49\textwidth]{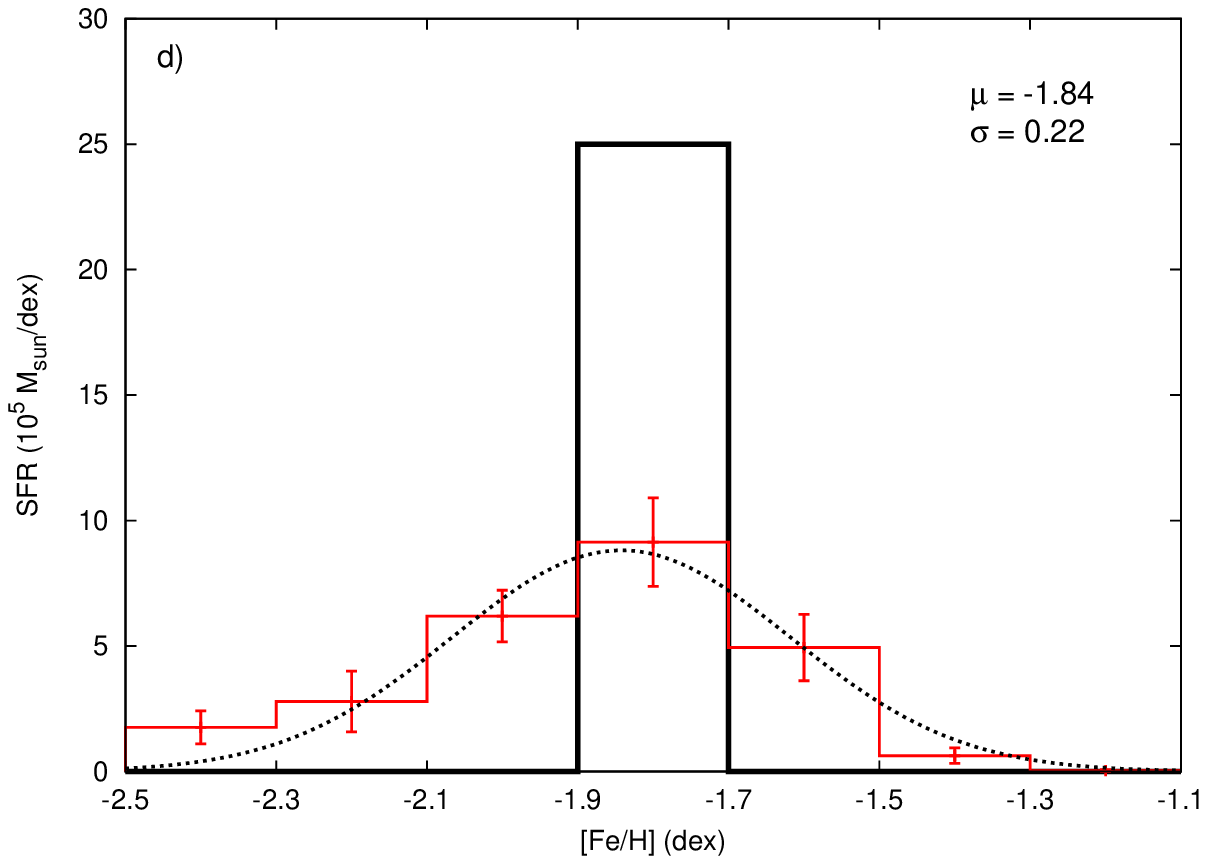}
\caption{\textbf{Upper row:} (\textbf{a}) Synthetic SFH and~(\textbf{b}) CEH of a single burst of star formation without adding any photometric errors. The input SFH is the solid black histogram, while the recovered SFH is the green histogram when using the same parameter gridding as the input. The red histogram gives the solution when using a set of different parameter griddings. \textbf{Bottom row:} A burst of star formation with realistic errors from artificial star tests. The black histogram in (\textbf{c}) the SFH and~(\textbf{d}) the ceh is the input SFH and the red histograms indicate the recovered SFH using a set of different population gridding schemes. A Gaussian fit to the recovered SFH is shown as a black line, with the mean~($\mu$) and variance~($\sigma$) also listed. \label{1burst}} 
\end{figure*}
\\
We determine the ability of Talos to recover the age, metallicity and SFR in a series of synthetic episodes of star formation. The stellar population was generated assuming the distance modulus and reddening of Sculptor, and using the artificial star tests to obtain realistic photometric uncertainties. In this way realistic colour magnitude diagrams can be obtained, of the same quality as the observed CMD of the Sculptor dSph. \\
As a first check, we apply the SFH fitting method to a single synthetic episode of star formation. The Dartmouth isochrones are used to generate a synthetic CMD with a continuous star formation between 9 and 10 Gyr and $-$1.9$<$[Fe/H]$<$$-$1.7 dex. First, the episode was generated without any photometric errors and the SFH determined using the exact parameter gridding of [Fe/H] and age as used to generate the population~(green histogram). The SFH is also determined using a set of different parameter griddings~(red histogram, see Figure~\ref{1burst}a,b) in order to test the effect of using different griddings to obtain the uncertainties on the SFH~(see Section~\ref{uncertainties}).  \\
Given exactly the same parameter gridding as the input population, the SFH is recovered at the right age and metallicity with the correct strength. A limited amount of ``bleeding" is observed, due to the uncertainties induced by the quality of the data. The effect of using a set of different griddings is more substantial bleeding of the star formation rate into neighbouring bins in age and metallicity~(as seen in the red histograms in Figure~\ref{1burst}a,b). This bleeding is a direct consequence of the quality of the photometric data and the use of different parameter griddings, and determines the SFH resolution~(see Section~\ref{uncertainties}). Since we do not have a priori knowledge of the duration and metallicity of an episode of star formation in the observed data, the choice of adopting a single population gridding will lead to biases in the results. Therefore, we always consider a range of different parameter griddings in obtaining the final SFH. 
\begin{figure*}[!htb]
\centering
\includegraphics[angle=0, width=0.49\textwidth]{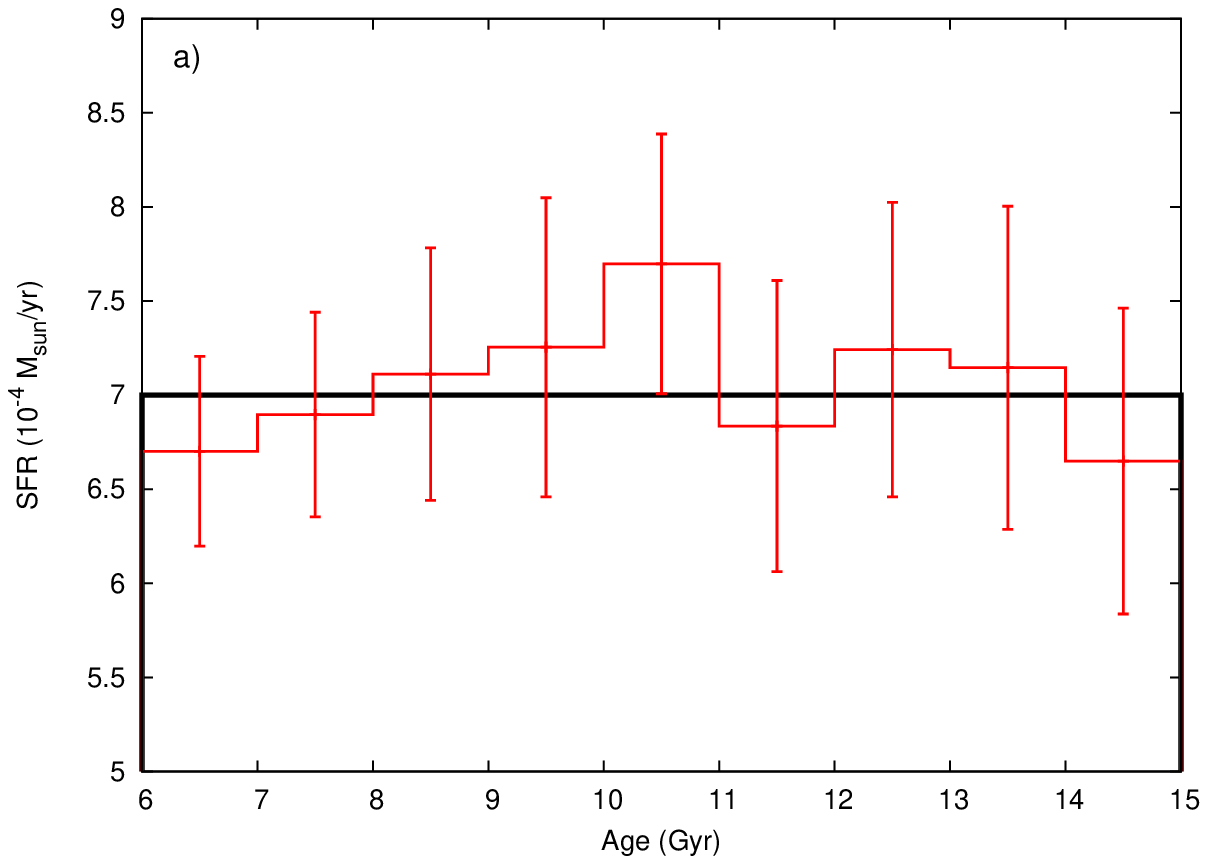}
\includegraphics[angle=0, width=0.49\textwidth]{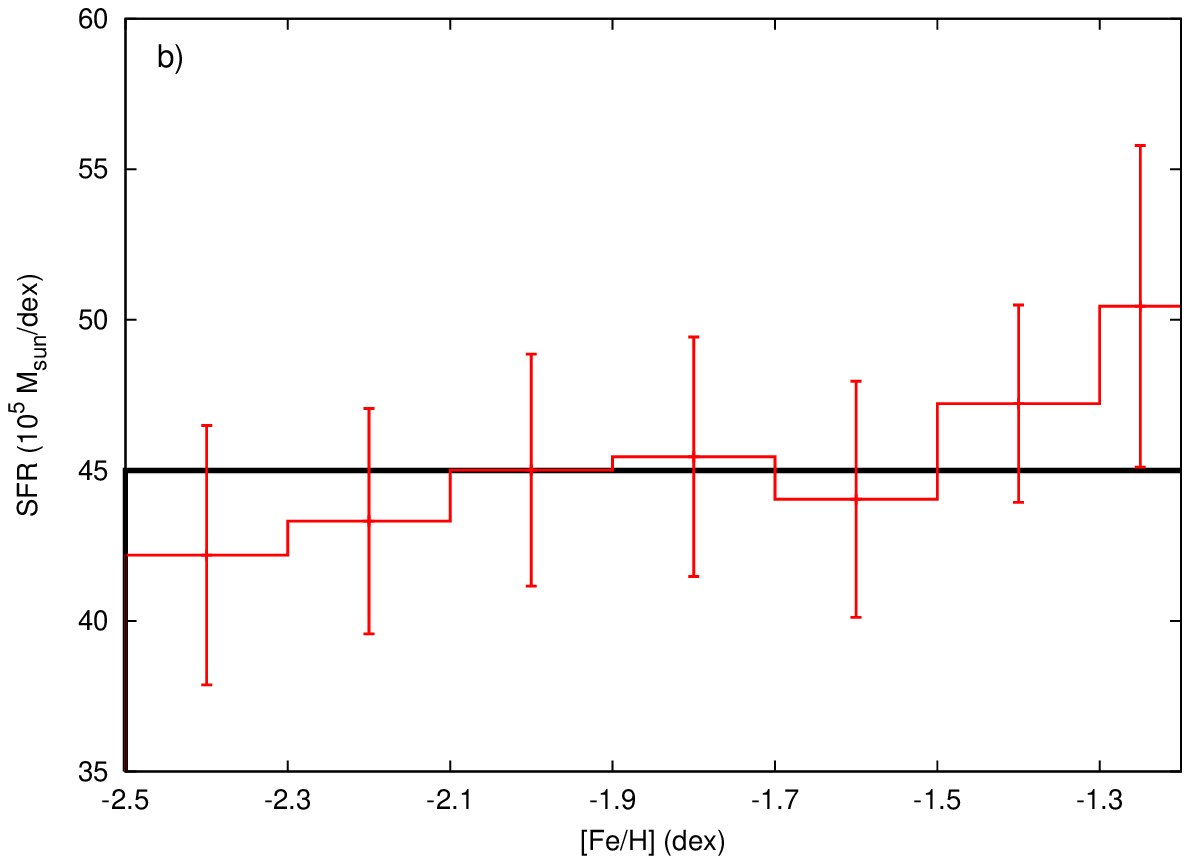}
\caption{(\textbf{a}) SFH and~(\textbf{b}) CEH of a continuous episode of star formation including realistic errors. The black line indicates the input values and the red histogram the recovered values. \label{contburst}} 
\end{figure*}
\\
When including realistic photometric errors determined from artificial star tests for the model of an episode of star formation, the same bleeding effect is seen. The parameters of the recovered SFH are determined by fitting a Gaussian profile, see Figures~\ref{1burst}c,d. The peaks of the SFH and CEH are recovered well within the input values ([Fe/H]$_{mean}$=$-$1.84 dex, Age$_{mean}$=9.64 Gyr), but the star formation is distributed over more bins, spreading out the star formation episode in time. Figure~\ref{1burst} shows that~$\approx$40\% of the total star formation is typically recovered within the central peak. Due to the quality of the observed photometric data~(which only just detects the oldest MSTOs) there remains some degeneracy between the burst strength and duration, which could be removed by obtaining deeper CMDs that resolve the MSTO with more accuracy. \\
Next we consider a more realistic synthetic population which has experienced constant star formation~(SFR=10$^{-4}$M$_{\odot}$/yr) over the metallicity range $-$2.5$<$[Fe/H]$<$$-$1.1 dex between 6 and 15 Gyr ago. To take into account the effect of constraints from spectroscopic observations, the solution was determined taking into account a synthetic MDF which samples 50\% of the RGB stars. The results are given in Figure~\ref{contburst}. It can be seen that the input values are correctly recovered for a synthetic population with constant star formation, within the SFH uncertainties.

\subsection{Globular cluster NGC 1904}
\begin{figure*}[!htb]
\centering
\includegraphics[angle=0, width=0.49\textwidth]{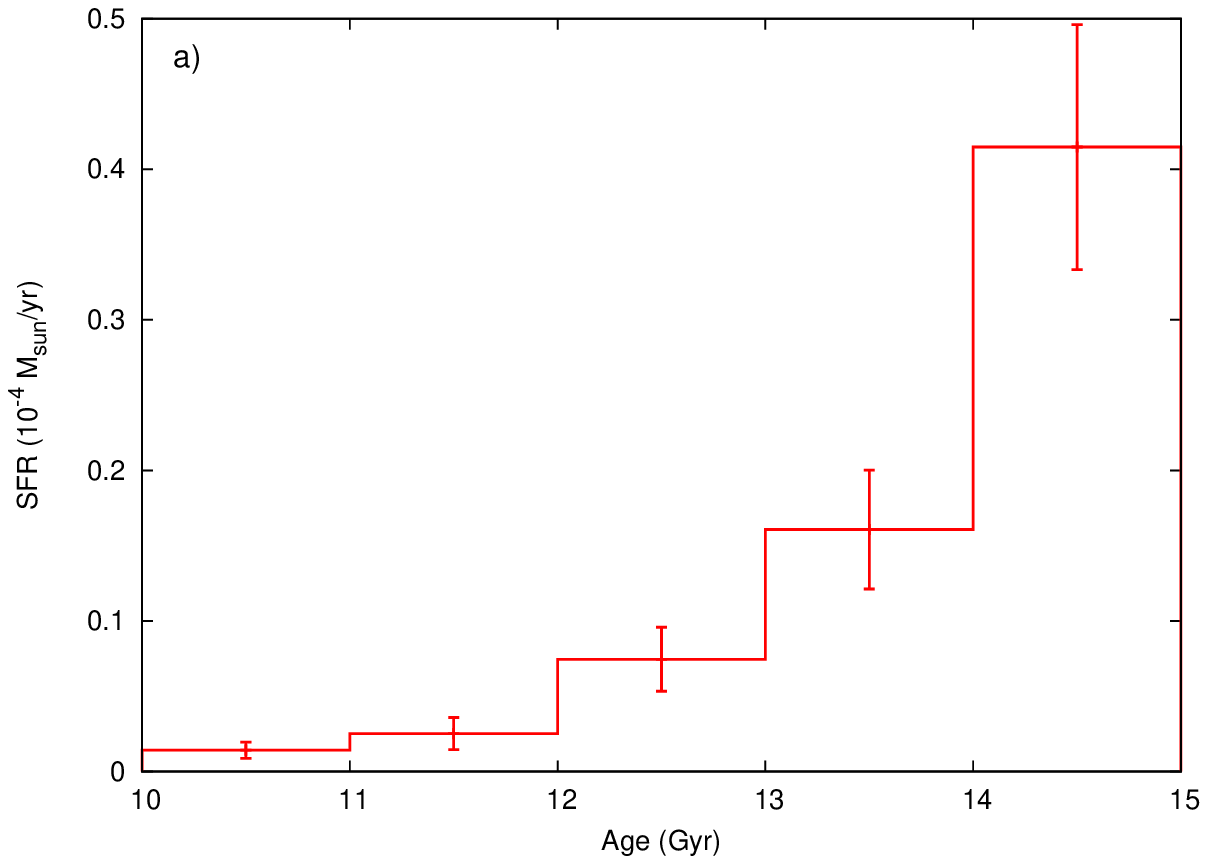}
\includegraphics[angle=0, width=0.49\textwidth]{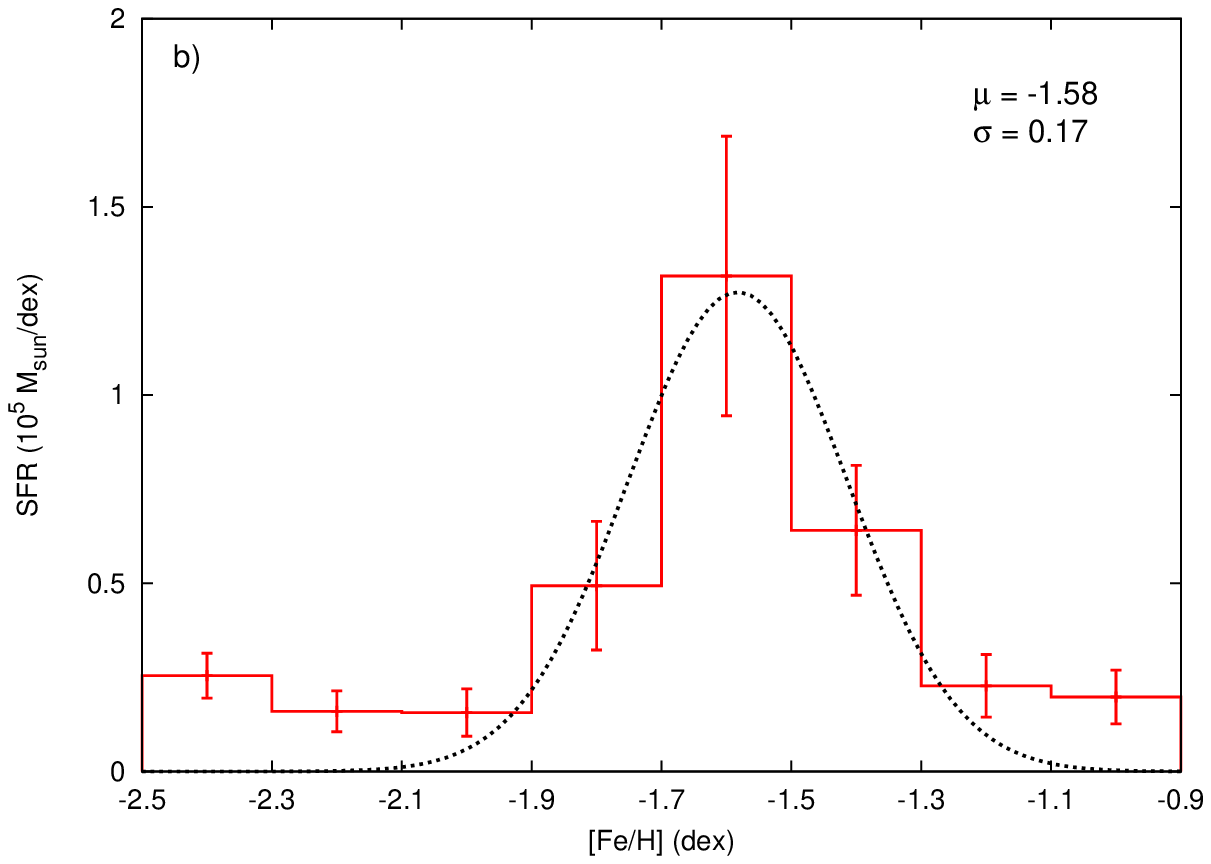}
\caption{(\textbf{a})~The SFH and~(\textbf{b}) CEH obtained for NGC1904 from photometric information in the B,V and I filters. A Gaussian fit to the MDF is also shown as a black, dotted line, with parameters given in the plot. \label{N1904SFH}} 
\end{figure*}

The final test we carry out is to apply our method to real observations of a Globular cluster, which is, within our errors, a simple stellar population. During our observing run Galactic globular cluster NGC 1904 was also observed in the B,V and I filters. For NGC1904 there have been several photometric and spectroscopic studies~(see Table~\ref{N1904pars}), making it a good test of our method. The reduction of these observations, as well as the artificial star tests were done in exactly the same way as for Sculptor. 
\begin{table}[!ht]
\caption[]{Adopted properties of NGC1904 \label{N1904pars}}
\begin{center}
\begin{tabular}{ccc}
\hline\hline
Property & Value & Reference \\
\hline
(m-M)$_{V}$ & 15.45$\pm$0.02 & \citep{Ferraro92}	\\
E(B-V) & 0.01 & \citep{Harris96}	\\
$[$Fe/H$]$ & $-$1.579$\pm$0.069 dex & \citep{Carretta09} \\
$[\alpha$/Fe$]$ & 0.31 dex &  \citep{Carretta10} 	\\
\hline 
\end{tabular}
\end{center}
\end{table}
\\
To obtain the SFH for NGC1904, [$\alpha$/Fe] was chosen as a fixed value constrained by spectroscopic observations~(see Table~\ref{N1904pars}), while the age and metallicity were left as free parameters. The best solution, using only the available photometric data is given in Figure~\ref{N1904SFH}. The SFH and CEH show bleeding into neighbouring bins, due to measurement errors and the use of different parameter griddings used. This is similar to the bleeding effect seen in Figure~\ref{1burst}. The SFH clearly indicates a very old population, in good agreement with typical globular ages. A Gaussian fit to the metallicity distribution gives a mean value~($\mu$) at [Fe/H]=$-$1.58 dex, with a variance of $\sigma$=0.17 dex. This is in good agreement with the spectroscopic [Fe/H] given in Table~\ref{N1904pars}. This shows that Talos is able to recover the age and metallicity of a real data set with all the errors and uncertainties that it implies. \\
These experiments show the capability of Talos to recover the age and metallicity of a stellar population, as well as the limit of our ability to unambiguously distinguish a burst of star formation from a more continuous SFR over a longer time. 

\end{appendix}

\end{document}